\documentclass[bibyear]{aa}  

\usepackage{graphicx}
\usepackage{hyperref}
\hypersetup{hidelinks}
\usepackage[dvipsnames]{xcolor}
\usepackage{marvosym}
\usepackage{natbib}
\bibpunct{(}{)}{;}{a}{}{,} 

\usepackage{subcaption,booktabs}
\usepackage{comment}
\usepackage[thinc]{esdiff}
\newcommand{\msun}{{\rm M}_\odot}
\newcommand{\fastcluster}{\textsc{fastcluster}$\,$}
\usepackage{soul}

\defcitealias{sana2012}{S12}
\defcitealias{sb2013}{SB13}
\usepackage{txfonts}
\usepackage{pifont}
\newcommand{\cmark}{\ding{51}}%
\newcommand{\xmark}{\ding{55}}%
\makeatletter
\renewcommand*\aa@pageof{, page \thepage{} of \pageref*{LastPage}}
\makeatother

\newcommand{\orcidicon}[1]{\href{https://orcid.org/#1}{\includegraphics[width=11pt]{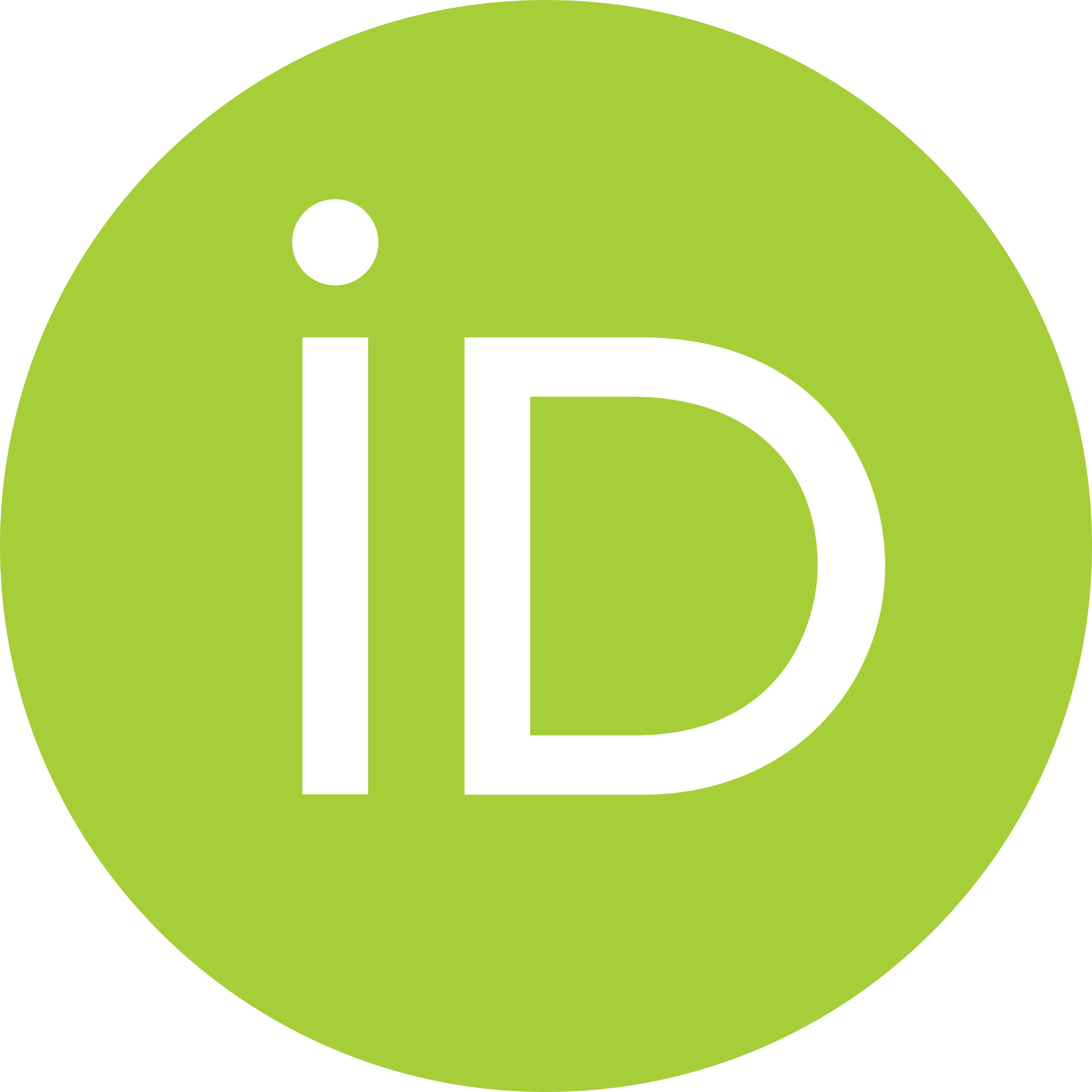}}}
\newcommand{\orcid}[1]{\href{https://orcid.org/#1}{\protect\orcidicon{#1}}}

\definecolor{seagreen}{rgb}{0.190, 0.525, 0.361}

\begin{document}

   \title{Binary black hole mergers from Population III star clusters}

   \author{Benedetta Mestichelli
          \inst{1,2,5}
    \orcid{0009-0002-1705-4729} \thanks{\href{mailto:benedetta.mestichelli@gssi.it}{benedetta.mestichelli@gssi.it}}
          \and 
          Michela Mapelli\inst{2,3,4} \orcid{0000-0001-8799-2548}
    \thanks{\href{mailto:mapelli@uni-heidelberg.de}{mapelli@uni-heidelberg.de}}
          \and
          Stefano Torniamenti\inst{2}
          \orcid{0000-0002-9499-1022}\thanks{\href{mailto:stefano.torniamenti@uni-heidelberg.de}{stefano.torniamenti@uni-heidelberg.de}}
          \and
          Manuel Arca Sedda\inst{1,5}
          \and \\
          Marica Branchesi\inst{1,5}
          \and
          Guglielmo Costa\inst{6}
          \and 
          Giuliano Iorio\inst{3,4}
          \and 
          Filippo Santoliquido\inst{1,5}
          }
          \authorrunning{B. Mestichelli et al.}
          \institute{
          $^1$Gran Sasso Science Institute (GSSI), Viale Francesco Crispi 7, 67100, L’Aquila, Italy\\
          $^2$Institut f{\"u}r Theoretische Astrophysik, ZAH, Universit{\"a}t Heidelberg, Albert-Ueberle-Stra{\ss}e 2, D-69120, Heidelberg, Germany\\
          $^3$Physics and Astronomy Department Galileo Galilei, University of Padova, Vicolo dell'Osservatorio 3, I--35122, Padova, Italy\\
          $^4$INFN - Padova, Via Marzolo 8, I--35131 Padova, Italy\\
          $^5$INFN, Laboratori Nazionali del Gran Sasso, I-67100 Assergi, Italy\\
          $^6$Univ Lyon, Univ Lyon1, ENS de Lyon, CNRS, Centre de Recherche Astrophysique de Lyon UMR5574, F-69230 Saint-Genis-Laval, France.
          }
    
   \date{}

 
  \abstract{Binary black holes (BBHs) born from the evolution of Population~III (Pop.~III) stars are one of the main high-redshift targets for next-generation ground-based gravitational-wave (GW) detectors. Their predicted initial mass function and lack of metals make them the ideal progenitors of black holes above the upper edge of the pair-instability mass gap, i.e. with a mass higher than $\approx{}134$ (241)~M$_\odot$ for stars that become (do not become) chemically homogeneous during their evolution. Here, we investigate the effects of cluster dynamics on the mass function of 
  BBHs born from Pop.~III stars, by considering the main uncertainties on Pop.~III star mass function, orbital properties of binary systems, star cluster's mass and disruption time. In our dynamical models, at least $\sim$5\% and up to 100\% BBH mergers in Pop.~III star clusters  have primary mass $m_1$ above the upper edge of the pair-instability mass gap. In contrast, only $\lesssim {} 3$\% isolated BBH mergers have primary mass above the gap, unless their progenitors evolved as chemically homogeneous stars. 
  The lack of systems with primary and/or secondary mass inside the gap defines a zone of avoidance with sharp boundaries in the primary mass -- mass ratio plane. 
  Finally, we estimate the merger rate density of BBHs and, in the most optimistic case, we find a maximum of $\mathcal{R}\approx200\,{\rm Gpc^{-3}\,yr^{-1}}$ at $z\sim15$ for BBHs formed via dynamical capture. For comparison, the merger rate density of isolated Pop.~III BBHs is  $\mathcal{R}\leq{}10\,{\rm Gpc^{-3}\,yr^{-1}}$, for the same model of Pop.~III star formation history.} 


   \maketitle
%

\section{Introduction}
The first generation of stars, called Population III (Pop.~III) stars, formed from zero-metallicity gas \citep{haiman1996, tegmark1997, yoshida2003} and most likely their initial mass function (IMF) included a larger contribution from  massive stars compared to stellar populations in the local Universe \citep{sb2013,susa2014,hirano2015,jaacks2019,sharda2020,liu2020,liu2020b,wollenberg2020,chon2021,tanikawa2021,jaura2022,prole2022}. Since Pop.~III stars lack in metals, they lose mass through stellar winds less efficiently; consequently, the mass of their compact remnants can be close to their zero-age main sequence (ZAMS) mass \citep{heger2002,kinugawa2014,kinugawa2016,hartwig2016,belczynski2017,tanikawa2021,tanikawa2021b,tanikawa2022,tanikawa2023,costa2023,santoliquido2023,tanikawa2024}. 
Thus, we may expect Pop.~III stars to efficiently produce intermediate-mass black holes (IMBHs), i.e. black holes (BHs) with mass in excess of $\approx{}100$~M$_\odot$ \citep{madau2001,woosley2002,liu2020b,kinugawa2021,tanikawa2021,tanikawa2021b,tanikawa2022,tanikawa2023,tanikawa2024}. However, the exact shape of the Pop.~III IMF 
is still largely unknown. The star formation history of these stars is poorly constrained as well. In the $\Lambda$CDM model, Pop.~III stars begin forming at $z\geq30$, but reach a peak in their star formation rate at $z\sim15-20$ (for a recent review on Pop.~III star formation, see \citealp{klessen2023}). 

Recent binary evolution simulations show that very few IMBHs born from Pop.~III~stars merge with other (IM)BHs \citep{tanikawa2021,tanikawa2021b,tanikawa2022,tanikawa2023,costa2023,tanikawa2024}. The main reason is that very-massive  Pop.~III~stars reach large radii ($>100$~R$_\odot$) during their life and thus either they collide prematurely with their companion star before becoming BHs, or they produce loose binary black holes (BBHs), whose initial semi-major axis is too wide to merge within the lifetime of the Universe.

However, if some of the Pop.~III stars formed in stellar clusters,  we expect stellar dynamics to efficiently couple and harden these massive BBHs \citep{heggie1975}. Hence, star-cluster dynamics might significantly boost the BBH merger rate from Pop.~III stars \citep{wang2022}. 
The first star clusters were embedded in dark matter  halos of $\sim10^7\,\msun$ \citep{reed2007, sakurai2017}, which could host star clusters with $M_{\rm cl}\lesssim10^5\,\msun$ \citep{bromm2002, sakurai2017, wang2022}. 
$N-$body models indicate that the presence of such dark matter halos can prevent the star cluster from disruption and keep it bound long enough to dynamically form and harden BBHs \citep{wang2022}. 

Current gravitational-wave (GW) interferometers are able to detect BBH mergers up to a redshift $z\sim{2}$ \citep{abbott2020,gwtc3_2023,abbott_pop2023}. Thus, they can detect BBH mergers from Pop.~III stars only when these happen at low redshift, with a long delay time. We expect such events to be outnumbered by BBH mergers originating from Population~I and II stars at low redshift \citep{santoliquido2023}. However,  starting from the next decade, third-generation GW detectors, such as Cosmic Explorer \citep{ce_2019} and Einstein Telescope \citep{et_2010} will have a much larger instrumental horizon, observing BBH mergers up to a redshift of $\sim{100}$ 
\citep{kalogera2021,branchesi2023}. Mergers of BBHs from Pop.~III stars will be one of the most wanted sources of GWs in the high-redshift Universe \citep{kalogera2021,ng2021,ng2022a,ng2022b,perigois2021,santoliquido2023,santoliquido2024}.

In this work, we explore the hierarchical mergers of Pop.~III BBHs in dense star clusters, while probing a large range of initial conditions for both their IMFs and formation channels of first-generation BBHs. We investigate the impact of the dynamical environment on the build-up and merger of IMBHs. 
To this purpose, we employ the semi-analytic code \fastcluster, which is conceived to model hierarchical BBH mergers in star clusters, while efficiently exploring the parameter space that this process involves. 
The paper is structured as follows. Section~\ref{methods} presents the initial conditions and outlines the numerical code. In Sec.~\ref{results}, we describe our main results. Sec.~\ref{discussion} discusses the main implications of our results, including the BBH merger rate density, and Sec.~\ref{conclusions} summarizes our conclusions.

\section{Methods} \label{methods}
In this work, we use the semi-analytic code \fastcluster \citep{mapelli2021, mapelli2022, vaccaro2023,torniamenti2024}, which integrates the effect of BBH dynamical hardening and GW emission, while capturing the main physical processes that drive the evolution of the host star cluster.
A complete description of  \fastcluster  can be found in \cite{mapelli2021,mapelli2022} and \cite{torniamenti2024}. In the following, we summarize the details of the set-up and initial conditions considered for this work.

\subsection{Pop.~III star models}\label{subsec:popIII}
We generated our first-generation (both single and binary) BHs  with the binary population synthesis code \textsc{sevn} \citep{spera2019,mapelli2020,sevn2023}, which calculates star evolution  by interpolating a set of pre-computed single stellar tracks 
and models the main binary interaction processes by means of semi-analytic prescriptions \citep{hurley2002, sevn2023}.

We computed the Pop.~III star tracks with the stellar-evolution code \textsc{parsec} \citep{bressan2012,chen2015,costa2019}. Specifically, we distinguish two different evolutionary paths of Pop.~III stars: "default" and chemically homogeneous evolution (CHE). We evolved the default tracks by assuming initially non-spinning stars with an initial hydrogen abundance $X=0.751$, helium abundance $Y=0.2485$ \citep{komatsu2011}, metallicity $Z=10^{-11}$ \citep{tanikawa2021,costa2023}, and considering the mass range between 2 and 600~M$_\odot$. 

For the CHE stars, we simulated a sample of pure-He tracks considering initial masses ranging from 0.36 to 350 M$_\odot$, with initial hydrogen abundance $X=0$, helium abundance $Y=1-Z$, and metallicity $Z=10^{-6}$. This metallicity is similar to the metal content we find in He cores of Pop.~III stars after the main-sequence phase. We ran these tracks as a proxy for CHE stars, which are deemed to be the result of fast-spinning metal-poor stars that become fully mixed already near the end of the main sequence \citep{mandel2016, demink2016,marchant2016, riley2022}. In the context of Pop.~III stars, it is important to explore the possibility of rapidly rotating stars, since simulations show that most of them have a rotational velocity between $50\%$ and $100\%$ of their Keplerian velocity \citep{sb2013}.  As our pure-He stars, CHE stars remain pretty compact during their entire evolution \citep{demink2016}. Since in our simplified treatment of CHE stars we skip the integration of the hydrogen main sequence, we then increase the BBH formation time by a factor equal to the duration of the main sequence.
We refer to Appendix~\ref{ap:che} and to \cite{costa2023} and \cite{santoliquido2023} for more details about our Pop.~III stellar tracks. 


The population synthesis code \textsc{sevn} assigns  to each star its final fate, based on a formalism for core-collapse and (pulsational) pair-instability supernovae. Namely, we adopted the rapid model for core-collapse supernovae \citep{fryer2012}, which enforces a mass gap between neutron stars and BHs, and the fitting formulas by \cite{mapelli2020} for (pulsational) pair-instability supernovae. Our model for pair instability results in a mass gap of the BH mass function between $\sim{86}$ and 241 M$_\odot$ for the default single-star models (see  Fig.~6 of \citealt{costa2023}). For the CHE models, the pair-instability mass gap is between $\sim{46}$ and $\sim{134}$ M$_\odot$, because after the main sequence  
these models evolve without H-rich envelope (see Fig.~17 of \citealt{santoliquido2023}).

For each of our samples (default and CHE), we used \textsc{sevn} in both single-star mode and binary-star mode. The single-star set-up allows us to generate single BHs for the dynamical BBH sample (see Section~\ref{subsec:dyn_bin}), while  the binary-star mode was used to produce our isolated and original BBHs (see Section~\ref{subsec:orig_bin}).

In the single-star set up, we have run six simulations with \textsc{sevn}.
Given the uncertainties on the  IMF of Pop.~III stars, we have run each model with a Kroupa \citep{kroupa2001}, a Larson \citep{larson1998}, a top-heavy (e.g. \citealp{liu2020}), and a log-flat (e.g. \citealp{jaura2022,hartwig2022}) IMF. We refer to \cite{costa2023} and to our Appendix~\ref{ap:isolated} for a description of these IMFs.

For the binary-star mode, we did not run new \textsc{sevn} models, but rather re-used the outputs of the simulations presented by \cite{costa2023}. These simulations assume a variety of initial conditions that encompass the uncertainties on Pop.~III binary star systems. Namely, they consider the same IMFs as we already summarized for the single star runs. For the mass ratios, they consider a uniform sampling, the distribution by \cite{sana2012} resulting from fitting formulas to nearby massive binary stars, or the results by \cite{sb2013}. 
For the initial orbital period distribution, they assume either the fit by \cite{sana2012} or the results by \cite{sb2013}. Finally, for the initial orbital eccentricities, they adopt either the fit by \cite{sana2012} or a thermal eccentricity distribution. We describe the set up of each run in our Appendix~\ref{ap:isolated}.

In the following sections, we assume the model \textbf{log1} \citep{costa2023} as our fiducial model. This assumes a log-flat IMF for the primary star mass  (e.g. \citealp{jaura2022,hartwig2022}), and the distributions by \cite{sana2012} for the mass ratio, orbital period, and eccentricity. In Sec.\ref{subsec:res_ic}, we show results found adopting other initial conditions. 

In their paper, \cite{costa2023} describe only the sub-sample of BBHs that merge over a timescale shorter than the lifetime of the Universe (measured by the Hubble time, $t_{\rm H}=13.6$~Gyr). Here, we call them \emph{isolated} BBH mergers: BBHs that form from isolated binary stars that are not members of stellar clusters and evolve unperturbed until they merge within $t_{\rm H}=13.6$ Gyr. We further describe the isolated BBH sample in Appendix~\ref{ap:isolated}. Here, we show the isolated BBHs only for the sake of comparison.

In addition, here we consider all the BBHs that form in the simulations by \cite{costa2023}, including those that have a delay time much longer than the Hubble time. In fact, star-cluster dynamics can either ionize or harden such BBHs, depending on whether they are soft or hard binary systems \citep{heggie1975}. Thus, some of the wide BBH binaries in \cite{costa2023} are likely to harden dynamically and merge within a Hubble time. In Appendix~\ref{ap:isolated}, we provide additional details about them.

\subsection{First-generation BBHs}
We consider two channels for the first-generation BBHs in star clusters: \emph{dynamical} (hereafter, Dyn) BBHs and \emph{original} (hereafter, Orig) BBHs. The former originate from single Pop.~III stars and then pair up dynamically inside their host star cluster, whereas the latter are BBHs that form from the evolution of  
original binary systems of Pop.~III stars.  Here, we call original binary systems those binary systems that are already present in the initial conditions of our star cluster simulations and then evolve dynamically inside the cluster. In stellar dynamics, these are usually referred to as primordial binaries \citep{goodman1989}, but here we use the term original binaries to avoid confusion with primordial BHs \citep{carr2016}. 

\subsubsection{Original BBHs} \label{subsec:orig_bin}

Original BBHs come from the binary-star simulations presented by \cite{costa2023}, who evolved several sets of Pop.~III binary stars with the \textsc{sevn} code. As described in Section~\ref{subsec:popIII}, our original BBHs consist of all the BBHs simulated by \cite{costa2023}, including those systems that are too wide to merge within a Hubble time. 

\subsubsection{Dynamical BBHs} \label{subsec:dyn_bin}
Dynamical BBHs form via three-body interactions at the time of core-collapse of the star cluster, when the central density of the cluster is the highest \citep{torniamenti2024}. 
In the case of dynamical BBHs, we draw the masses of the primary BHs out of the \textsc{sevn} single BH catalogs, i.e. we assume that these BHs come from the evolution of single stars. We assign the primary and secondary BH masses using two different pairing functions. 
The first one, which we will call DynA 
samples $m_1$ uniformly, while $m_2$ follows the probability distribution $p(m_2)\propto{}(m_1+m_2)^4$, as found by  \cite{oleary2016}. As shown in \cite{torniamenti2024}, this sampling criterion can encompass the uncertainties connected with the BH pairing function, e.g. due to stochastic fluctuations, stellar and binary evolution, BH-star interactions.

The second pairing function, hereafter DynB, 
follows the mass pairing criterion adopted by \cite{antonini2023} and based on \cite{heggie1975}. This function describes the dynamical formation of hard BBHs through three-body encounters, and systematically leads to the pairing of the most massive objects. For further details, see Appendix~\ref{ap:pairing}.

We sample the initial orbital eccentricity and semi-major axis of the dynamical BBHs from the distributions
\begin{equation} \label{eq:ecc_peri}
\begin{aligned}
    p(e) &= 2e,\,\,\,\,\,\,\,\,\mathrm{with}\,\,e\in[0,1), \\
    p(a) &\propto a^{-1},\,\,\,\,\,\,\,\,\mathrm{with}\,\,a\in[1,10^3]\,\mathrm{R_{\odot}}.
\end{aligned}
\end{equation}
The above distributions are commonly used for hard binary systems in star clusters \citep{mapelli2021}.

\subsubsection{Spins of first-generation BBHs}

For both original and dynamical BBHs, we generate the dimensionless spin magnitudes $\chi_1$ and $\chi_2$ from a Maxwellian distribution with $\sigma_{\chi}\,=\,0.05$ and truncated at $\chi=1$. This choice for the spin magnitudes is a toy model matching the properties of LIGO--Virgo BBHs by construction \citep{abbottO3popandrate}. The spin directions are drawn isotropically over a sphere, since dynamical interactions remove any alignment of the spins with the angular momentum of the binary system \citep{rodriguez2016}.

\subsubsection{Additional checks for first-generation BBHs}
For both original and dynamical BBHs, we check whether the binary is hard, i.e. if the binding energy of the BBH is larger than the average kinetic energy of a star:
\begin{equation}
    E_{\rm b} = \frac{G\,m_1\,m_2}{2\,a}\,>\,\langle E_{\rm k}\rangle=\frac{1}{2}\,m_*\,\sigma^2,
\end{equation}
with $m_1$ and $m_2$ the masses of primary and secondary stars at ZAMS (BHs) in the case of original (dynamical) BBHs, $m_*$ the average mass of a star in the cluster (depending on the chosen IMF) and $\sigma$ the three-dimensional velocity dispersion. In this work, we neglect dynamical exchanges, which might enhance the mass of the BBH components. On the other hand, the effect of dynamical exchanges is already taken into account in the BBH formation rate of model DynB (see Appendix~\ref{ap:pairing}). If the binary is soft, it is not integrated further, because we assume that it is disrupted by dynamical interactions. 

After drawing the masses, we check whether the BHs are expelled from the cluster by a supernova kick. We estimate the natal kick as previously done by \cite{mapelli2021},
\begin{equation}
    v_{\rm SN} = v_{\rm H05}\frac{\langle m_{\rm NS}\rangle}{m_{\rm BH}}
\end{equation}
with $\langle m_{\rm NS}\rangle\,=\,1.33\,\msun$ the average neutron star mass \citep{ozel2016}, $m_{\rm BH}$ the BH mass and $v_{\rm H05}$ a number drawn from a Maxwellian distribution with $\sigma_{\rm v}\,=\,265\,\mathrm{km\,s^{-1}}$. This quantity is compared to the escape velocity of the cluster that we compute as \citep{georgiev2009}:
\begin{equation}\label{eq:vesc}
    v_{\rm esc}=40\,\mathrm{km\,s^{-1}}\,\left(\frac{M_{\rm cl}}{10^5\,\msun}\right)^{1/3}\left(\frac{\rho}{10^5\,\msun\,\mathrm{pc^{-3}}}\right)^{1/6},
\end{equation}
where $M_{\rm cl}$ is the mass of the cluster, and $\rho$ is the density at the half-mass radius. If the primary or secondary mass in a dynamical BBH have $v_{\rm SN}>v_{\rm esc}$, the binary is no longer integrated. In the case of an original BBH, the orbital evolution of the binary is computed outside the cluster. 

\subsection{Orbital evolution and N-th generation BBHs}
We evolve the semi-major axis $a$ and eccentricity $e$ using the formalism presented by \cite{mapelli2021}, where both the dynamical hardening \citep{heggie1975} and  GW decay \citep{peters1964} are taken into account:
\begin{equation}
\begin{aligned}
\diff{a}{t} &= -2\,\pi\,\xi\,\frac{G\,\rho_{\rm c}}{\sigma}a^2-\frac{64}{5}\frac{G^3\,m_1\,m_2\,(m_1+m_2)}{c^5\,a^3(1-e^2)^{7/2}}f_1(e),\\
\diff{e}{t} &= 2\,\pi\,\xi\,\kappa\,\frac{G\,\rho_{\rm c}}{\sigma}a-\frac{304}{15}e\frac{G^3\,m_1\,m_2\,(m_1+m_2)}{c^5\,a^4(1-e^2)^{5/2}}f_2(e),
\end{aligned}
\end{equation}
where $\rho_{\rm c}$ is the core density, $\xi$ is a numerically calibrated constant  that we fix to $\xi=3$ \citep{hills1983, quinlan1996, sesana2006}, and $\kappa={\rm d}{e}/{\rm d}\ln{(1/a)}$. 
The functions $f_1(e)$ and $f_2(e)$ are defined by \cite{peters1964} as
\begin{equation}
\begin{aligned}
f_1(e) &=\left(1+\frac{73}{24}e^2+\frac{37}{96}e^4 \right), \\
f_2(e) &=\left(1+\frac{121}{304}e^2\right).
\end{aligned}
\end{equation}

If a BBH merges within the lifetime of its cluster (see Sec.~\ref{subsec:clust_prop}), we compute the spin and mass of the remnant  using the fitting formulas by \cite{jimenez2017}. Moreover, we estimate the magnitude of the relativistic kick $v_{\rm k}$ experienced by the remnant with eq. 12 from \cite{lousto2012}. If $v_{\rm k}<v_{\rm esc}$ (with $v_{\rm esc}$ from eq. \ref{eq:vesc}) the merger remnant remains inside the cluster and can pair up again, otherwise it is ejected and we do not consider it anymore. 

We generate the nth-generation secondary BHs using the pairing criteria described in Sec. \ref{subsec:dyn_bin} and Appendix \ref{ap:pairing}. Here, we assume that the secondary BH is always first-generation and $m_2\in\left[m_{\rm min},\max(m_{\rm 1,1g})\right]$ with $\max(m_{\rm 1,1g})$ the maximum mass of a first-generation primary BH. \cite{torniamenti2024} also consider the case in which the secondary BH can be a nth-generation BH and find no substantial differences.

We repeat the hierarchical merger process until one of the following conditions is reached: the merger remnant is ejected from the star cluster, the cluster evaporates, no BHs are left inside the cluster, or we reach $z_{\rm min}$ (Sec. \ref{subsec:clust_prop}).

\subsection{Ejection by three-body encounters}\label{sec:3bejection}

It is essential to know whether a BBH is ejected from its parent cluster as a consequence of a binary-single star scatter before it can merge. If this happens, the BBH will merge outside the cluster and the final stages of its evolution will be driven only by gravitational-wave decay, without dynamical hardening. In \fastcluster{}, we use a simple argument based on energy exchanges during binary-single star encounters to estimate this process \citep{miller2002}. 
Specifically, we expect that the gravitational recoil induced by a binary-single encounter can lead to the ejection of a BBH if its semi-major axis is shorter than \citep{miller2002}
\begin{equation}
    a_{\mathrm{ej}} = \frac{2\,\xi\,m_*^2}{(m_1+m_2)^3}\frac{G\,m_1\,m_2}{v_{\rm esc}^2}.
\end{equation}

For comparison, we estimate the maximum semi-major axis for which the shrinking by emission of GWs starts dominating over the shrinking by dynamical hardening \citep{baibhav2020}
\begin{equation}
    a_{\mathrm{GW}} = \left[\frac{32\,G^2}{5\,\pi\,\xi\,c^5}\frac{\sigma\, m_1\,m_2\,(m_1+m_2)}{\rho_{\rm c}\,(1-e^2)^{7/2}}\,f_1(e)\right]^{1/5}.
\end{equation}
Statistically, we expect that the BBH will merge inside the cluster  when $a_{\rm GW}>a_{\rm ej}$, which is the criterion we adopt in \fastcluster{} (these mergers will be labeled as \textit{InCl}). Otherwise, we assume that the BBH is ejected and merges in the field (label \textit{Ej3B}).

\subsection{Properties of the star clusters} \label{subsec:clust_prop}
Cosmological hydrodynamic simulations \citep{sakurai2017} indicate that Pop.~III star clusters tend to form in dark matter halos with $m_{\rm DM}\sim10^7\,\msun$. These star clusters form at epochs up to $z\sim20$, and present typical masses $M_{\rm cl}\lesssim10^5\,\msun$ \citep{sakurai2017,wang2022}. At the same time, magneto-hydrodynamic simulations \citep{chong2019} show that, under conditions that reproduce the ones expected in high-redshift galaxies, molecular clouds up to $\sim 10^5\,\msun$ produce a majority of star clusters with $\sim10^4\,\msun$ . 

In this work, we consider both high-mass (HM)  and low-mass (LM) star clusters,  assuming that they both form at $z=20$. We draw the initial cluster masses from a log-normal distribution with mean $\langle\log_{10}{M_{\rm cl}/\msun}\rangle\,=\,5.6$ and 4.3  for HM and LM clusters, respectively, and standard deviation $\sigma_M\,=\,0.4$. We draw their density at half-mass radius from a log-normal distribution with mean $\langle\log_{10}{\rho/(\mathrm{M_{\odot}\,pc^{-3}})}\rangle\,=\,3.7$ and 3.3 for HM and LM clusters respectively, and $\sigma_\rho\,=\,0.4$. This choice of parameters for the HM clusters is inspired by \cite{sakurai2017} and \cite{wang2022}, whereas for the LM clusters we adopted typical values for low-redshift young massive clusters \citep{mapelli2022}. We will consider a cosmologically-motivated mass distribution in a follow-up work.

We integrate star cluster evolution following the model in \cite{torniamenti2024}, based on the formalism presented in \cite{antonini2020}. 
In this work, we do not consider mass loss from tidal stripping from the host galaxy but we evolve the clusters until a minimum redshift $z_{\rm min}$, at which we assume them to be disrupted. After the disruption, the BBHs evolve and eventually merge in isolation. The results presented in the following were obtained taking $z_{\rm min}=2$; in Sec.~\ref{subsec:disruption_gc}, we compare them to results found with $z_{\rm min}=10,\,6$.

\subsection{Merger rate density}
We estimate the BBH merger rate density 
using the semi-analytic code \textsc{cosmo}$\mathcal{R}$\textsc{ate} \citep{santoliquido2020,santoliquido2021}, which combines catalogs of simulated BBH mergers with a chosen star formation rate density model. Specifically, the merger rate density in a comoving frame is computed as:
\begin{equation}
\label{eq:mrd}
    \mathcal{R}=\int_{z_{\rm max}}^z\psi(z')\diff{t(z')}{z'}\left[\int_{Z_{\rm min}}^{Z_{\rm max}}\eta(Z)\mathcal{F}(z',z,Z)\,\mathrm{d}Z\right]\,\mathrm{d}z'.
\end{equation}
Here, $\psi(z')$ is the chosen star formation rate density; we choose 
the star formation history derived by the semi-analytic model of \cite{hartwig2022}, which traces
individual Pop.~III stars, and is calibrated on several observables from the local and high-redshift Universe. In Eq.~\ref{eq:mrd}, $\mathrm{d}t(z')/\mathrm{d}z'=H_0^{-1}(1+z')^{-1}[(1+z')^3\,\Omega_M+\Omega_{\Lambda}]^{-1/2}$, with $H_0$ the Hubble parameter, and $\Omega_{M}$ and $\Omega_{\Lambda}$ the matter and dark energy densities \citep{aghanim2020}.
The merger efficiency $\eta(Z)$ is defined differently for original and dynamical BBHs. In the case of dynamical BBHs, we specify it as
\begin{equation} \label{eq:eta_dyn}
    \eta_{\rm Dyn}(Z) = \frac{\mathcal{N}_{\rm merge,sim}(Z)}{\mathcal{N}_{\rm sim}(Z)}\,\frac{\mathcal{N}_{\rm BH}(Z)}{M_{\rm cl,TOT}(Z)}.
\end{equation}
with $\mathcal{N}_{\rm merge,sim}(Z)$ the number of simulated BBH mergers and $\mathcal{N}_{\rm sim}(Z)$ the number of simulated BBHs. $\mathcal{N}_{\rm BH}(Z)$ is instead the number of BHs we expect to form from a stellar population, assuming a certain IMF and that all stars with ZAMS mass above $20\,\msun$ are BH progenitors. Finally, $M_{\rm cl,TOT}(Z)$ is the total initial mass of the simulated clusters. 

In the case of original BBHs, we define the merger efficiency as
\begin{equation}
    \eta_{\rm Orig}(Z)= \frac{\mathcal{N}_{\rm merge,sim}(Z)}{\mathcal{N}_{\rm sim}(Z)}\,\frac{\mathcal{N}_{\rm TOT}(Z)}{M_*(Z)},
\end{equation}
where the second ratio is a formation efficiency of the BBHs in our \textsc{sevn} catalogs, with $\mathcal{N}_{\rm TOT}(Z)$ the total number of BBHs and $M_*(Z)$ the total initial stellar mass.
The term $\mathcal{F}(z',z,Z)$ is given by
\begin{equation}
\label{eq:f_mrd}
    \mathcal{F}(z',z,Z)=f_{\rm{CL}}(z',M_{\rm{cl}})\frac{\mathcal{N}(z',z,Z)}{\mathcal{N}_{\rm TOT}(Z)}\,p(z',Z),
\end{equation}
where $f_{\rm{CL}}(z',M_{\rm{cl}})$ is the fraction of stellar mass contained in HM and LM clusters at redshift $z'$; we set $f_{\rm{CL}}=1$, assuming that all the stellar mass at $z'$ is either included in HM or LM clusters. $\mathcal{N}(z',z,Z)$ gives us the number of BBHs forming at redshift $z'$ with metallicity $Z$ and merging at redshift $z$. Lastly, since we fix $Z=10^{-11}$ ($Z=10^{-6}$) for the default (CHE) stars, $p(z',Z)$ is a delta function different from zero only at $Z=10^{-11}$ ($Z=10^{-6}$).

\section{Results} \label{results}

\begin{figure*}[ht]
    \centering
    \includegraphics[width=0.9\textwidth]{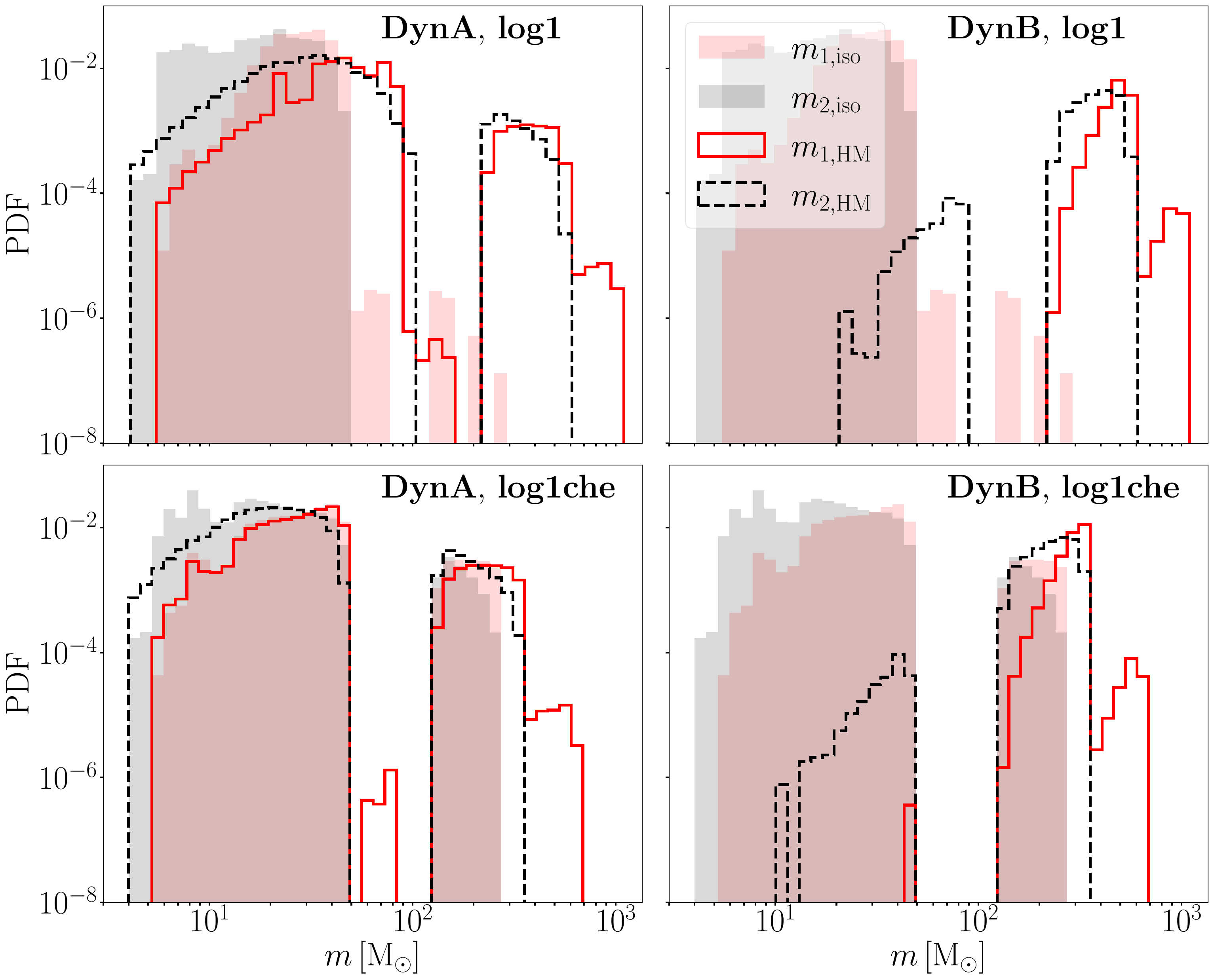}
    \caption{Distributions of primary (red) and secondary (black) masses of BBHs merging in HM clusters, for  \textbf{log1} (upper panels) and \textbf{log1che} (lower panels). The shaded histograms show the results found by \protect\cite{costa2023} for isolated BBHs. The left column shows the results found with the pairing function DynA, while the right one shows the results with DynB.}
    \label{fig:m1m2_gc}
\end{figure*}

\begin{table}
\centering
\centering
\begin{tabular}{c c c c c} 
 \hline
 IC & Isolated & Orig & DynA & DynB\\ [0.5ex]
    & \%       & \%   & \%   & \% \\
 \hline\hline
 \textbf{kro1} & $5.23\times10^{-4}$ & 2.15 & 5.33 & 99.99\\
 \textbf{kro5} & 0 & 60.63 & -- & --\\
 \textbf{lar1} & 0 & 2.67 & 5.85 & 99.99\\
 \textbf{lar5} & $4.32\times10^{-3}$ & 65.27 & -- & --\\
 \textbf{lar5che} & 11.17 & 52.66 & 10.88 & 99.99\\
 \textbf{log1} & $5.26\times10^{-4}$ & 11.48 & 35.15 & 100\\
 \textbf{log1che} & 39.29 & 42.42 & 45.93 & 99.99\\
 \textbf{log2} & 0 & 93.97 & -- & --\\
 \textbf{log3} & 3.05 & 17.51 & -- & --\\
 \textbf{log4} & 0 & 29.04 & -- & --\\
 \textbf{log5} & 0.02 & 94.49 & -- & --\\
 \textbf{top1} & 0 & 20.67 & 70.91 & 97.96\\
 \textbf{top5} & 0.04 & 97.96 & -- & --\\
\hline
\end{tabular}
\vspace*{1mm}
\caption{Percentage of BBH mergers with primary mass above the pair-instability mass gap. The first column reports the name of the models (see Appendix~\ref{ap:isolated} for details); the second the fraction of isolated BBH mergers with $m_1$ above the gap. The third, fourth and fifth show the same fraction in the case of original, DynA, and DynB BBH mergers inside HM clusters.}
\label{table:fraction_massive}
\end{table}

\subsection{Dynamical versus isolated BBHs}

Figure~\ref{fig:m1m2_gc} displays the primary and secondary mass distribution of  dynamical BBH mergers. For comparison, we also show the isolated BBHs from \cite{costa2023}. We consider both \textbf{log1} and \textbf{log1che} BBHs in HM clusters. 

Star-cluster dynamics boosts the merger rate of Pop.~III BBHs above the pair-instability mass gap by several orders of magnitude.
Actually, most BBH mergers have primary and secondary members above the mass gap if we adopt the pairing function by \cite{antonini2023}:  
model DynB (both \textbf{log1} and \textbf{log1che}) yields almost no BBH with $m_1<200\,\msun$. 

We find a dramatic difference between isolated BBHs and dynamical BBHs: BBH mergers above the gap are rare in  isolated BBHs, while they are 
common in dynamical BBHs.
This difference is maximal in the case of model DynB (both with and without CHE): BBH mergers with $m_1$ above the mass gap are at least $\sim{98}$\% of the dynamical BBHs (both with and without CHE), while they are much less common in isolated BBHs (at most $\sim{39}$\% for CHE models). Model DynA, on the other hand, yields a percentage of BBH mergers above the gap spanning from $\sim{5}\%$ to $\sim{71}\%$.


In the models with CHE  (e.g., \textbf{log1che}), we find less difference between isolated and dynamical BBHs above the gap compared to the default models. In fact, CHE favors the merger of isolated BBHs above the mass gap, since the stellar radii are compact through the entire stellar evolution (see the discussion by \citealt{santoliquido2023}). This allows very massive stars to evolve in a tight binary system without colliding prematurely, so that they end up producing massive tight BBHs.  

Even in the CHE models, star-cluster dynamics allows the pair-up and merger of more massive BBHs and the formation of hierarchical chains that push $m_1$ up to $\sim700\,\msun$. As above, the pairing criterion DynB definitely favors the formation and merger of BBHs with $m_1$ above the upper mass gap. 

In Table \ref{table:fraction_massive}, we present the fraction of BBH mergers with $m_1$ above the pair-instability mass gap for all the simulated models. We see in general how dynamical BBHs (especially those paired with DynB) involve a primary BH above the gap. Models with a larger initial orbital period from \cite{sb2013} (those ending with "5" and \textbf{log2}), produce little or no merging systems with $m_1$ above the upper mass gap in isolation; the effect of star-cluster dynamics is to efficiently harden these BBHs and lead to their merger.

\subsection{Original versus dynamical BBHs}\label{subsec:prim_dyn}

\begin{figure*}[ht]
    \centering
    \includegraphics[width=0.9\linewidth]{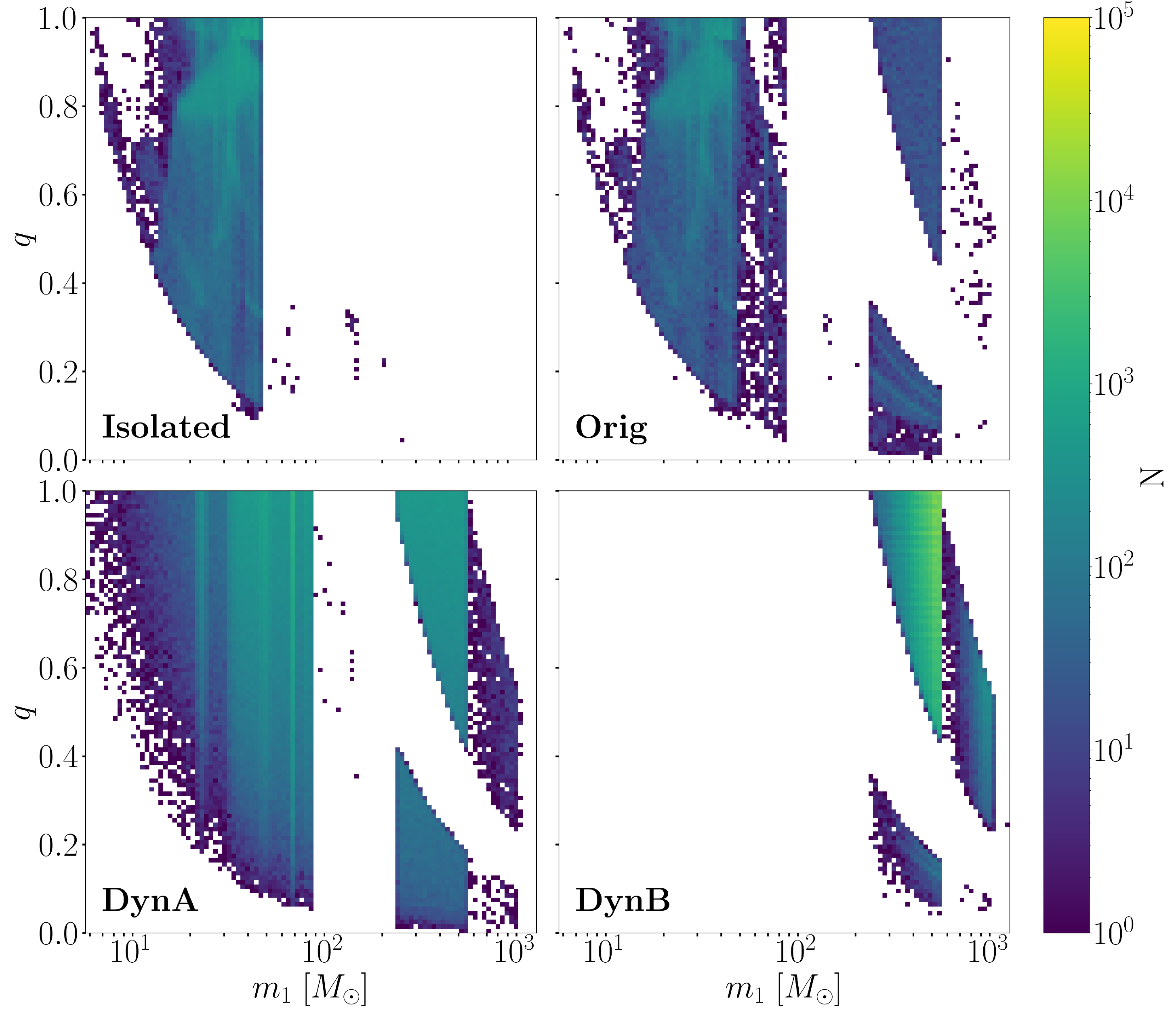}
    \caption{Distribution of the mass ratio $q$ as a function of the primary mass $m_1$ for model \textbf{log1}. We show the results for isolated, original (Orig) and dynamical (DynA and DynB) binaries in HM clusters. The color map shows the number of BBH mergers per cell.}
    \label{fig:m1q_log1}
\end{figure*}

Figure~\ref{fig:m1q_log1} shows the distribution of the mass ratio $q$ of BBH mergers as a function of their primary mass for the \textbf{log1} model. We show isolated binaries \citep{costa2023}, original binaries, and dynamical binaries (for both pairing functions).

Overall, the distribution of $m_1-q$ for the original binaries is similar to that of the isolated binaries, but for one key difference: the dynamical hardening in stellar clusters boosts the merger rate of original BBHs above the pair-instability mass gap compared to isolated BBHs. Above the gap, original binaries behave more like dynamical BBHs than like isolated BBHs. 

Figure~\ref{fig:m1q_log1} also shows that we expect several sharp features in the $m_1-q$ plot of both original and dynamical BBHs because of the pair-instability mass gap. Specifically, we expect a zone of avoidance in the $m_1-q$ plane, corresponding to the values for which $m_1$ is above the pair-instability mass gap and $m_2$ would be inside the mass gap. Such feature is not apparent for isolated Pop.~III BBH mergers because of the scarcity of mergers above the gap, unless we assume CHE.

\begin{figure*}[ht]
    \centering
    \includegraphics[width=0.9\linewidth]{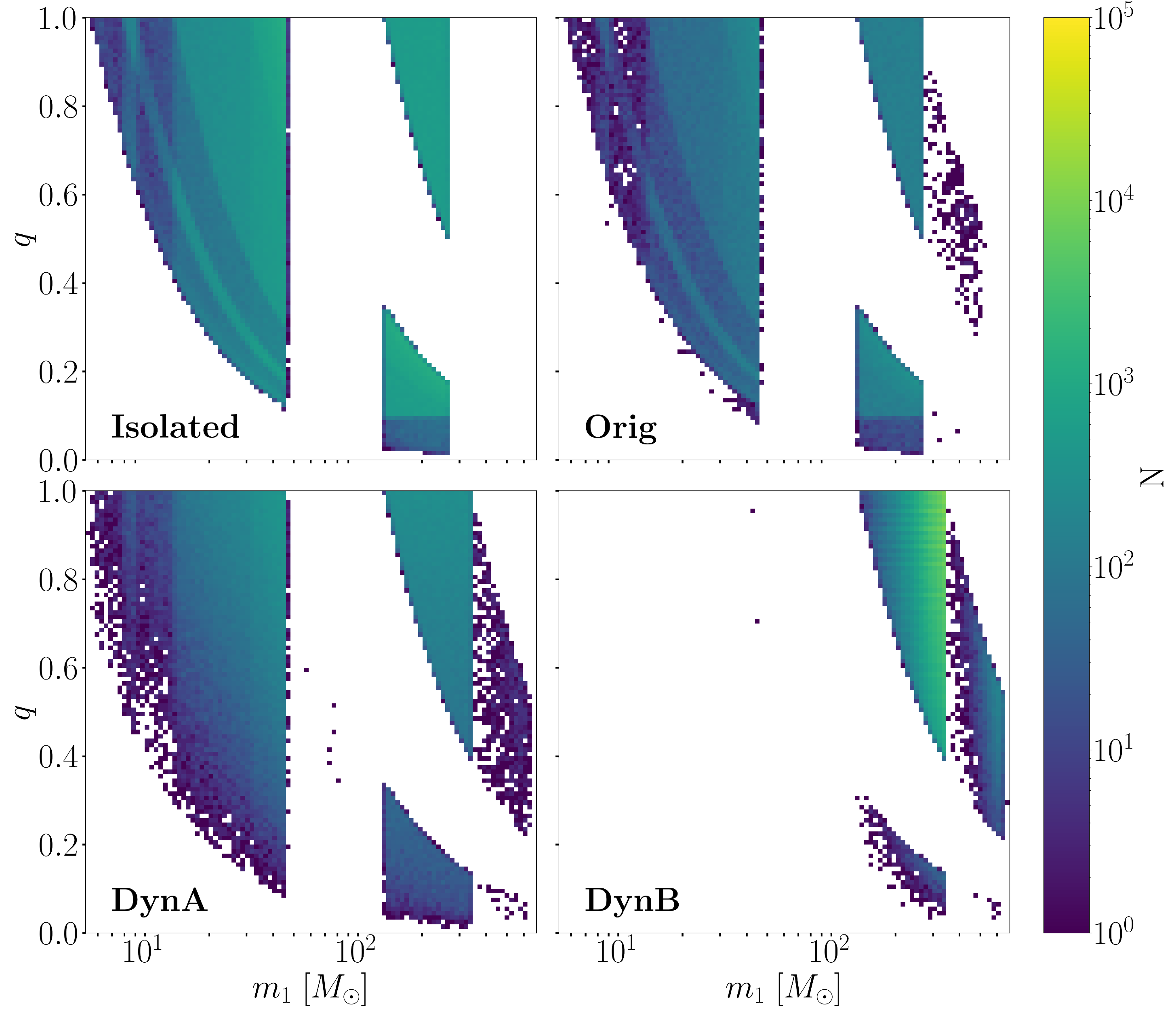}
    \caption{Same as Figure~\ref{fig:m1q_log1} but with \textbf{log1che}.}
    \label{fig:m1q_log1che}
\end{figure*}

Figure~\ref{fig:m1q_log1che} shows the same distributions for \textbf{log1che}. 
In the case of \textbf{log1che}, 
the distributions of isolated, original and dynamical BBHs in the $m_1-q$ are more similar to each other, because CHE enhances the merger rate above the mass gap even in isolated BBHs.

\subsection{Impact of cluster properties}\label{subsec:clust_effect}

\begin{figure*}[ht]
    \centering
    \includegraphics[width=0.9\textwidth]{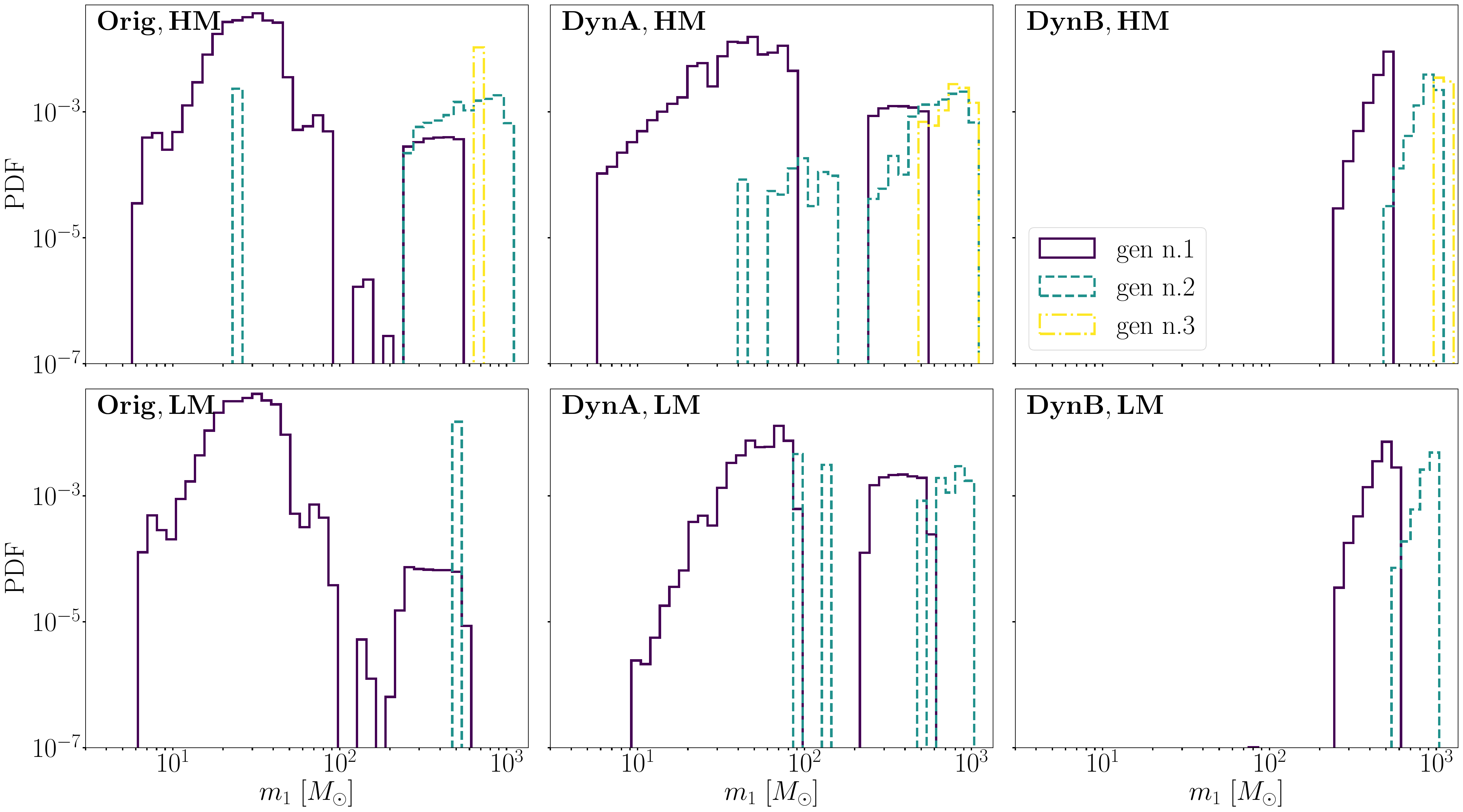}
    \caption{Distribution of the primary mass $m_1$ of BBH mergers at various generations both in HM (top row) and in LM clusters (bottom row). We show  model \textbf{log1}.}
    \label{fig:m1_ngen_log1}
\end{figure*}

Figure~\ref{fig:m1_ngen_log1}  shows the primary mass distribution of multiple BBH merger generations obtained by simulating HM and LM clusters with \textbf{log1}. We do not show the \textbf{log1che} model, because it yields very similar results.

The BBH mergers in LM clusters populate less the region above $m_1\sim600\,\msun$, and lesser systems generate a chain of mergers compared to HM clusters. 
The reason is that LM clusters are less massive and less dense than HM clusters; as a consequence, their typical escape velocity is lower ($v_{\rm esc}\sim15\,\mathrm{km\,s^{-1}}$). The effect of a lower escape velocity is twofold. First, it reduces the number of low-mass BHs that  remain inside the cluster and merge, because even small supernova kicks and three-body recoils are sufficient to eject them. 
Second, a lower escape velocity also quenches high-generation mergers, because the  relativistic kick is more easily above $v_{\rm esc}$, leading to the ejection of merger remnants. 
For example, in the case of \textbf{log1}, we observe 14\% less original BBH mergers, and 44\% less DynA BBH mergers in the LM clusters compared to the HM clusters.
Furthermore, $\gtrsim82\%$ less mergers happen inside the cluster in LM clusters with respect to HM clusters, mainly because of supernova kicks and three-body ejections. 

The maximum number of generations is three in HM and two in LM clusters. Models DynB reach the largest primary mass: $m_1\sim1200\,\msun$  and $m_1\sim1000\,\msun$ in HM and LM clusters, respectively. 
As expected, the upper mass gap gets partially populated by higher order mergers.

\subsection{Impact of Pop.~III star initial conditions}\label{subsec:res_ic}

\begin{figure*}[ht]
    \centering
    \includegraphics[width=\textwidth]{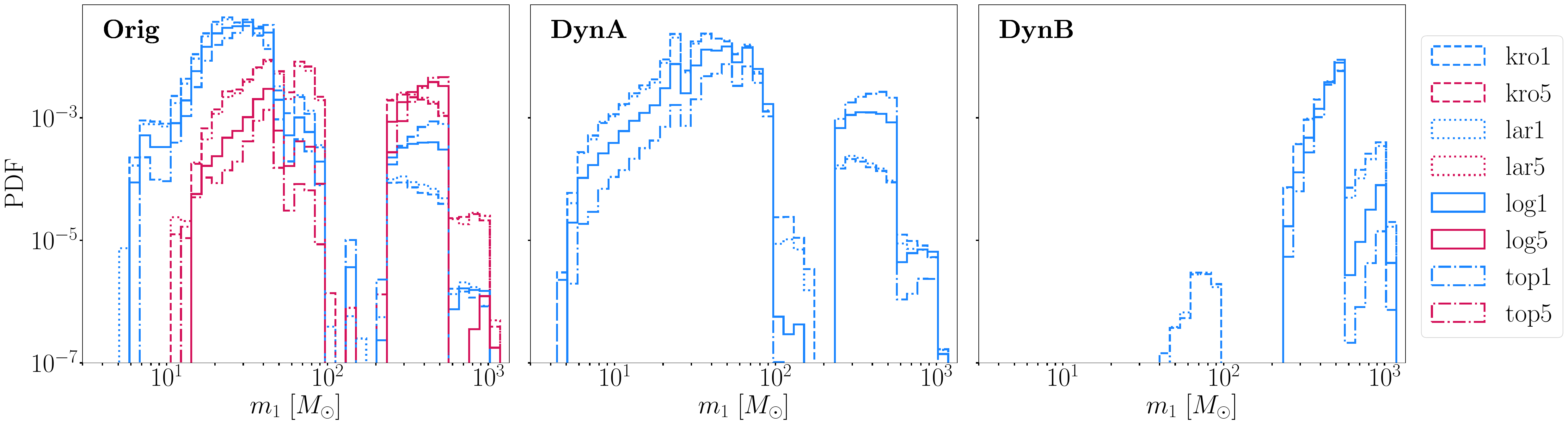}
    \caption{Primary mass distribution of BBH mergers for different initial conditions, defined in Appendix~\ref{ap:isolated} (Table~\ref{table:ic_costa}). The light-blue (red) lines show initial conditions that assume the orbital period distribution by  \citealp{sana2012} (\citealp{sb2013}) for the progenitor stars.}
    \label{fig:m1_ic}
\end{figure*}

\begin{figure*}[ht]
    \centering
    \includegraphics[width=\textwidth]{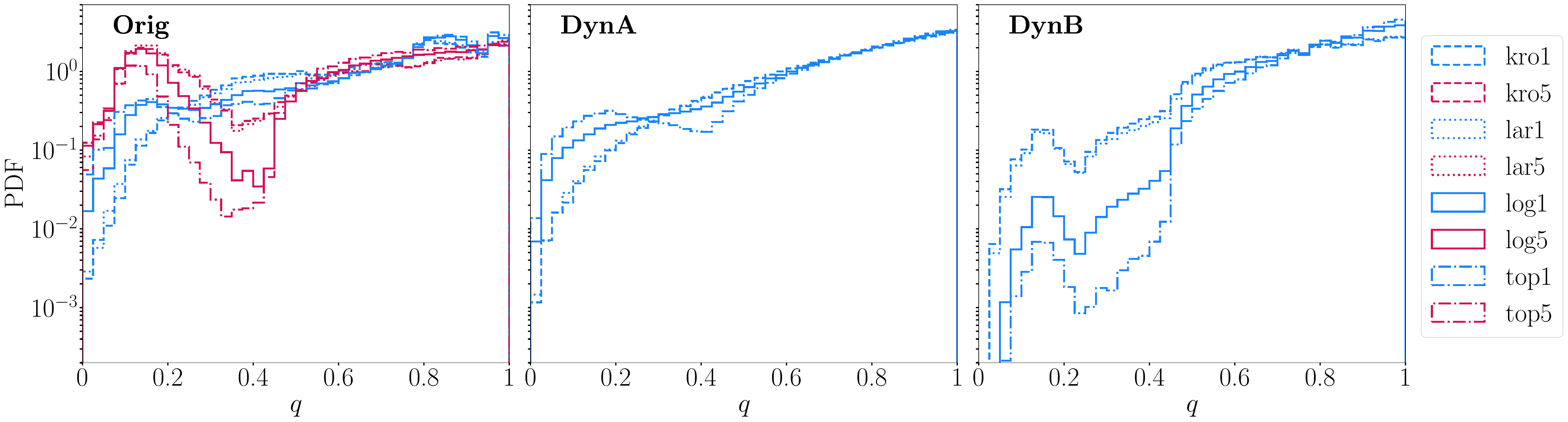}
    \caption{Mass-ratio distributions of BBH mergers for different initial conditions, defined in Appendix~\ref{ap:isolated} (Table~\ref{table:ic_costa}). We show in light-blue the initial conditions based on \cite{sana2012} and in red the ones based on \cite{sb2013}.}
    \label{fig:q_ic}
\end{figure*}

Figures~\ref{fig:m1_ic} and \ref{fig:q_ic} show the  primary mass and  mass ratio of BBH mergers for most of the initial conditions considered in \cite{costa2023} (see Appendix \ref{ap:isolated}). We do not show all of them to avoid that the figure becomes unreadable. 
Initial conditions labelled with "1" adopt an initial orbital period distribution from \cite{sana2012}, while initial conditions labelled with "5" follow the distribution by \cite{sb2013}. The former have shorter initial orbital periods than the latter, on average.

Figure~\ref{fig:m1_ic} shows that, while at $m_1<600\,\msun$ (first generation BBH mergers) the shape of the IMF directly influences the shape of the mass distributuon of BBH mergers
(see Sec.2.3 of \citealp{costa2023}), at higher generations the dynamical interactions inside the cluster play an extremely important role. For instance,  \textbf{top1} produces less BBHs with primary mass $m_1\sim10^3\,\msun$ with respect to an IMF peaking at lower masses such as \textbf{log1}. As a matter of fact, we find that the more the IMF is peaked at high masses, the more BBHs are ejected from the cluster because of three-body interactions. As we can see in Table \ref{table:flag_orig}, the percentage of BBH mergers ejected by three-body encounters (\textit{Ej3B}) increases as we move to IMFs peaking at higher and higher values.

On the other hand, CHE systems tend to merge inside the cluster more often but are also more prone to be ejected by relativistic kicks outside the cluster. The first property is most likely related to the presence of CHE binaries at high $e$, leading to larger values of $a_{\rm GW}$. The second one, could instead arise from the peak of BBH mergers at low mass ratios, more evident in the case of CHE systems.

\begin{table*}
\centering
\caption{Final fate of BBH mergers in HM clusters}
\begin{subtable}{\textwidth}
\caption{Fate of original BBH mergers}
\centering
\begin{tabular}{c c c c c c c} 
 \hline
 IC & Ej3B & EjSN & InCl & EjGR & Cl\Cross & AfterDisr\\ [0.5ex]
  & \% & \% & \% & \% & $\times10^{-3}$\% & $\times10^{-3}$\%\\
 \hline\hline
 \textbf{kro1} & 96.98 & 2.36 & 0.11 & 0.54 & 0 & 1.75\\ 
 \textbf{kro5} & 92.59 & 0.14 & 1.72 & 5.55 & 0 & 0\\
 \textbf{lar1} & 97.33 & 1.96 & 0.12 & 0.58 & 0.60 & 0.60\\ 
 \textbf{lar5} & 93.37 & 0.09 & 1.71 & 4.82 & 2 & 0 \\
 \textbf{lar5che} & 83.55 & 0 & 3.01 & 13.42 & 2.31 & 9.25\\
 \textbf{log1} & 98.36 & 1.33 & 0.12 & 0.19 & 0 & 0\\ 
 \textbf{log1che} & 97.75 & 1.51 & 0.18 & 0.56 & 0.45 & 0\\
 \textbf{log2} & 99.84 & 0.04 & 0.06 & 0.06 & 0 & 0\\
 \textbf{log3} & 98.23 & 1.22 & 0.48 & 0.07 & 2.65 & 0\\
 \textbf{log4} & 98.18 & 1.09 & 0.30 & 0.43 & 0 & 0\\
 \textbf{log5} & 99.84 & 0.04 & 0.06 & 0.06 & 0 & 0\\
 \textbf{top1} & 99.18 & 0.80 & 0.02 & 0.01 & 0 & 0\\ 
 \textbf{top5} & 99.96 & 0.01 & 0.01 & 0.02 & 0 & 0\\
\hline
\end{tabular}
\vspace*{1mm}
\caption*{\fastcluster flags describing the fate of original BBH mergers in HM clusters. The first column reports all the considered initial conditions, and the following columns contain the percentages of BBH mergers with a specific flag. \textit{Ej3B} means that the systems was ejected by three-body encounters; \textit{EjSN} stands for binary systems ejected from the cluster by a supernova kick; \textit{InCl} means that the BBH merger happens inside the cluster and that its remnant remains inside the cluster; \textit{EjGR} denotes systems merging inside the cluster whose remnant is kicked outside of it because of the relativistic kick; \textit{Cl\Cross} represents the BBHs merging outside the cluster after its evaporation; finally \textit{AfterDisr} describes the percentage of BBHs merging at $z<z_{\rm min}$, with $z_{\rm min}=2$.}
\label{table:flag_orig}
\end{subtable}
\vspace*{5mm}
\begin{subtable}{\textwidth}
\caption{Fate of dynamical BBH mergers (model DynA)}
\centering
\begin{tabular}{c c c c c c} 
  \hline
 IC & Ej3B & InCl & EjGR & Cl\Cross & AfterDisr\\ [0.5ex]
  & \% & \% & \% & $\times10^{-3}$\% & $\times10^{-3}$\%\\
 \hline\hline
 \textbf{kro1} & 95.27 & 1.03 & 3.69 & 0.62 & 5.42\\ 
 \textbf{lar1} & 97.09 & 0.68 & 2.23 & 0 & 1.02\\
 \textbf{lar5che} & 96.99 & 0.74 & 2.27 & 0 & 4.77\\
 \textbf{log1} & 98.68 & 0.42 & 0.90 & 0 & 0\\ 
 \textbf{log1che} & 98.71 & 0.39 & 0.89 & 0 & 0\\
 \textbf{top1} & 99.67 & 0.12 & 0.21 & 0 & 0\\ 
\hline
\end{tabular}
\vspace*{1mm}
\caption*{Same as Table \ref{table:flag_orig} but in the case of dynamical BBH mergers using the pairing criterion DynA. In this case no BBH merger can have the \textit{EjSN} flag, because we assume that the single BHs ejected by supernova kick will not pair up anymore.}
\label{table:flag_u}
\end{subtable}
\vspace*{5mm}
\begin{subtable}{\textwidth}
\caption{Flags of dynamical BBH mergers (model DynB)}
\centering
\begin{tabular}{c c c c c c} 
  \hline
 IC & Ej3B & InCl & EjGR & Cl\Cross & AfterDisr\\ [0.5ex]
 & \% & \% & \% & \% & \%\\
 \hline\hline
 \textbf{kro1} & 27.57 & 11.56 & 60.73 & 0.01 & 0.13\\ 
 \textbf{lar1} & 48.53 & 9.43 & 41.99 & $2.27\times10^{-3}$ & 0.04\\
 \textbf{lar5che} & 65.19 & 6.61 & 28.15 & 0 & 0.05\\
 \textbf{log1} & 94.39 & 1.74 & 3.87 & 0 & 0\\ 
 \textbf{log1che} & 95.79 & 1.28 & 2.93 & 0 & $5.51\times10^{-4}$\\
 \textbf{top1} & 99.30 & 0.28 & 0.42 & 0 & 0\\ 
\hline
\end{tabular}
\vspace*{1mm}
\caption*{Same as Table \ref{table:flag_u} but in the case of dynamical BBH mergers using the pairing function DynB.}
\label{table:flag_t}
\end{subtable}
\end{table*}

\subsection{Features in the mass ratio}\label{subsec:low_q}
Figure~\ref{fig:q_ic} shows the distribution of mass ratios  for both original and dynamical BBH mergers. 
Several features appear in the mass ratio of original BBHs and they are markedly different for models labeled with "1" (progenitors' orbital period from \citealt{sana2012}) and models labeled with "5" (progenitors' orbital period from \citealt{sb2013}).

In models assuming the initial orbital period from \cite{sb2013} (\textbf{kro5}, \textbf{lar5}, \textbf{log5}, \textbf{top5}), we find
\begin{enumerate}[(i)]
    \item a peak at $q\sim0.15$, 
    \item a drop at $q\sim0.4$, 
    \item a steady increase for $q>0.4$.
\end{enumerate}

As already discussed in Sec.~\ref{subsec:prim_dyn}, 
the peak for $q\sim0.15$ 
is mainly due to BBHs with $m_1$ above the upper mass gap and $m_2$ under the gap. The peak is particularly evident in the case of the orbital periods from  \cite{sb2013} because this set of initial conditions favors the formation of BBHs with low mass ratios \citep{costa2023}. The majority of the BBH mergers with $q\sim0.15$ in our simulations derive from binaries that did not undergo any kind of mass transfer; these binaries did not merge within a Hubble time as isolated binaries \citep{costa2023}. In original binaries, their merger is sped up by dynamical hardening.

In models assuming an initial orbital period from \cite{sana2012} (\textbf{kro1}, \textbf{lar1}, \textbf{log1}, \textbf{top1}), we find
\begin{enumerate}[(i)]
    \item a drop at $q\leq{}0.4$ (with the hint of a possible peak at 0.15 for models \textbf{log1} and \textbf{top1}),
    \item a steady increase for $q>0.4$,
    \item a peak at $q\sim0.8$. 
 \end{enumerate}

The peak at $q\sim0.8$ for models assuming the \cite{sana2012} orbital-period distribution is mainly due to BBHs with both $m_1$ and $m_2$ smaller than $\sim100\,\msun$ (Sec.\ref{subsec:prim_dyn}). As already discussed by   \cite{costa2023}, these systems form through stable mass transfer; the mass ratio is low enough that the first mass transfer happens while the secondary is still on the main sequence. 

The central and right plot of Figure~\ref{fig:q_ic} show the mass-ratio distributions of dynamical BBHs using either the DynA or the DynB pairing criterion. Here, we do not distinguish between models adopting \cite{sana2012} or \cite{sb2013}, because these BBHs form from stars that were initially \emph{single}.
Both DynA and DynB models have a strong preference for nearly equal-mass systems. However, we also see a peak at $q\sim{0.15}$, corresponding to BBHs with primary mass above the gap and secondary mass below the gap. 
In models DynB, we also find a steeper drop at $q$ below 0.4--0.5 because such models yield a strong preference for mergers with both $m_1$ and $m_2$ above the mass gap (Fig.~\ref{fig:m1m2_gc}).

\section{Discussion} \label{discussion}

\subsection{Tidal disruption of GCs}\label{subsec:disruption_gc}

We simulated our star clusters without taking into account the tidal field of their host galaxy. Hence, we do not know a priori when they will get disrupted. We will include this calculation in a forthcoming study. Here, we simply want to quantify how important is our assumption about the time at which the parent star cluster gets disrupted. Hence, we assume that the star cluster is completely disrupted at redshift $z_{\rm min}= 10$, 6 and 2, respectively. In Figure~\ref{fig:m1q_z_log1} (Appendix~\ref{ap:tidal_gc}), we show that the mass distribution of BBH mergers does not vary for $z_{\rm min}<10$. The variation is $<0.028\%$ for \textbf{log1} between $z_{\rm min}=10$ and $z_{\rm min}=2$. 

The formation time of dynamical BBHs mainly depends on the dynamical friction timescale  \citep{chandra1943}
\begin{equation}
    t_{\rm DF} = \frac{3}{4\,(2\pi)^{1/2}\,G^2\,\ln{\Lambda}}\,\frac{\sigma^3}{m_{\rm BH}\,\rho}
\end{equation}
with $m_{\rm BH}$ the mass of the black hole segregating at the center of the cluster, $\rho$ the density at half-mass radius, $\sigma$ the 3D velocity dispersion, and $\ln{\Lambda}$ the Coulomb logarithm. The more massive the BH is, the less time it requires to sink at the center of the cluster and pair up with another BH. As a consequence, all the massive BHs in our clusters readily form a chain of mergers by $z=10$.

In a recent work by \cite{wang2022}, the authors hypothesized that the presence itself of a dark matter halo around the star cluster would protect it from tidal disruption. The assumption of a Navarro, Frenk and White \citep{nfw1996} potential for the dark matter halo allows the cluster to survive down to our redshift. 
However, the lack of observations of Pop.~III clusters in the local Universe leads us to believe that such systems were disrupted either during or shortly after the re-ionization epoch.

The fact that the clusters got disrupted at high redshift does not mean that we cannot observe Pop.~III BBH mergers in the low-redshift Universe. To get a sense of what we can observe at low redshift with LIGO--Virgo--KAGRA, we keep integrating our Pop.~III BBHs down to redshift zero, assuming that they evolve only via gravitational-wave decay after the disruption of their parent cluster.
As we see from Fig.~\ref{fig:m1q_z_log1}, the mass function of BBH mergers does not depend on the $z_{\rm min}$ of disruption of the cluster.

\subsection{Merger rate density}
\begin{figure}[ht]
    \centering
    \includegraphics[width=\columnwidth]{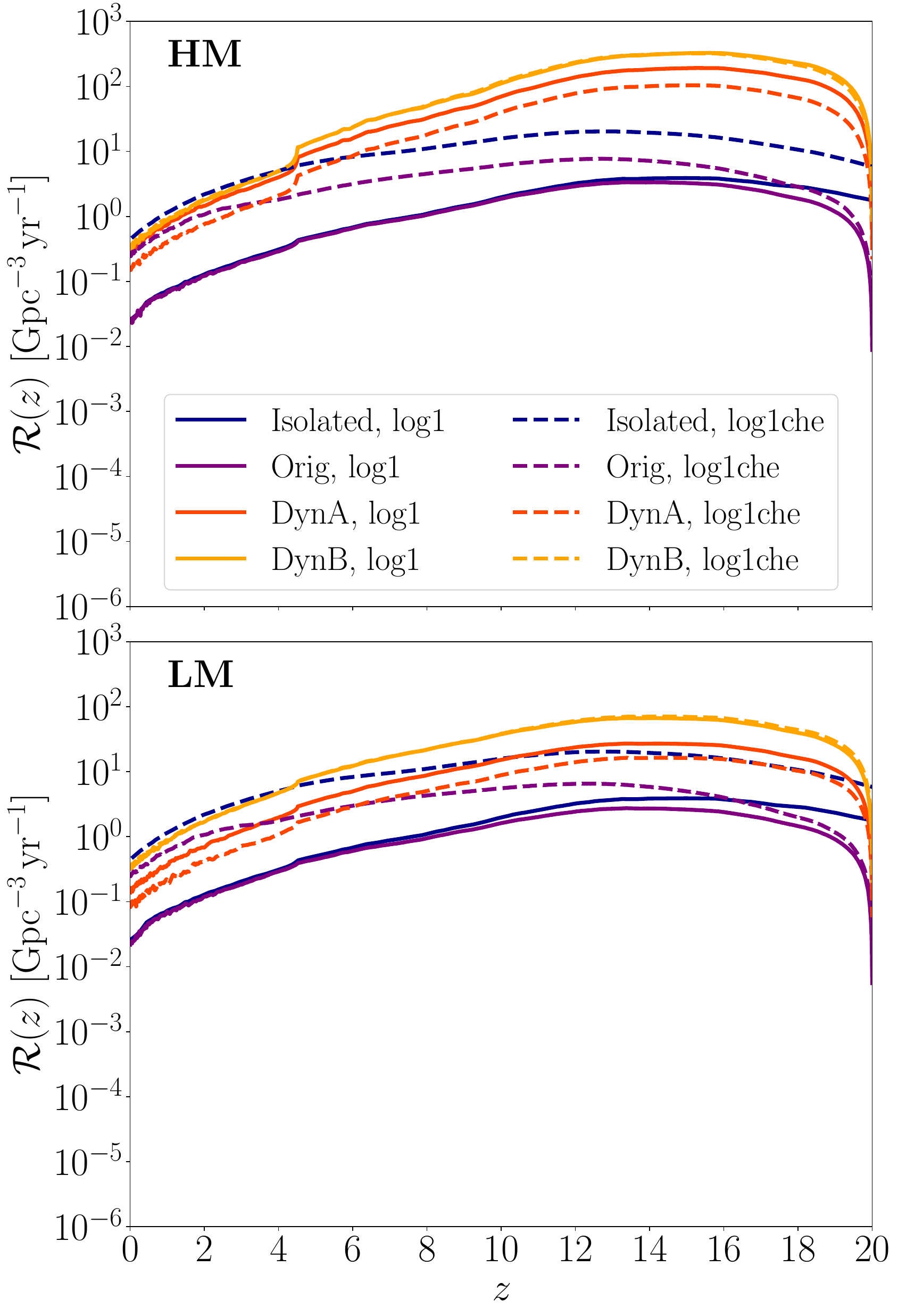}
    \caption{Merger rate density of Pop.~III BBHs in star clusters.  We assume that all of our clusters are disrupted at $z_{\rm min}=10$. We show both \textbf{log1} (solid lines) and \textbf{log1che} (dashed lines). Isolated and original BBHs are shown in blue and purple, respectively. DynA and DynB BBHs are shown in red and orange, respectively. On the upper (lower) plot, we show the BBH merger rate density in HM (LM) clusters.}
    \label{fig:mrd_gc_ysc}
\end{figure}

Figure~\ref{fig:mrd_gc_ysc} shows the BBH merger rate density in HM and LM clusters. We present our results for $z_{\rm min}=10$ only, since the mass distributions of BBH mergers do not vary at lower redshifts.


The merger rate density of dynamical BBHs (in HM and LM clusters) is higher than the one of both original and isolated BBHs, especially at high redshift ($z\gtrsim{}5$). Also, the merger rate density of dynamical BBHs peaks at $z\sim15$ (earlier than both original and isolated BBHs) and then rapidly decreases. In fact, most dynamical BBHs merge rapidly, with a short delay time. At later times we can only observe the least massive BBH mergers, characterized by longer delay times.

In general, the merger rate density of BBHs in LM clusters is lower than the one in HM clusters, because of a lower merger efficiency (see Sec.~\ref{subsec:clust_effect}). In fact, LM clusters lose their black holes more efficiently than HM clusters.

BBHs born from CHE Pop.~III stars yield a higher merger rate density both in the case of isolated and original BBHs. This feature was already discussed in \cite{santoliquido2023}, and is due to the higher merger efficiency of CHE BBHs.  Also, the merger rate density of isolated BBHs tends to be higher than that of original BBHs in star clusters. The reason is that star-cluster dynamics tends to break some of the original binary stars, that is the ones that are initially soft.

\begin{figure}[ht]
    \centering
    \includegraphics[width=\columnwidth]{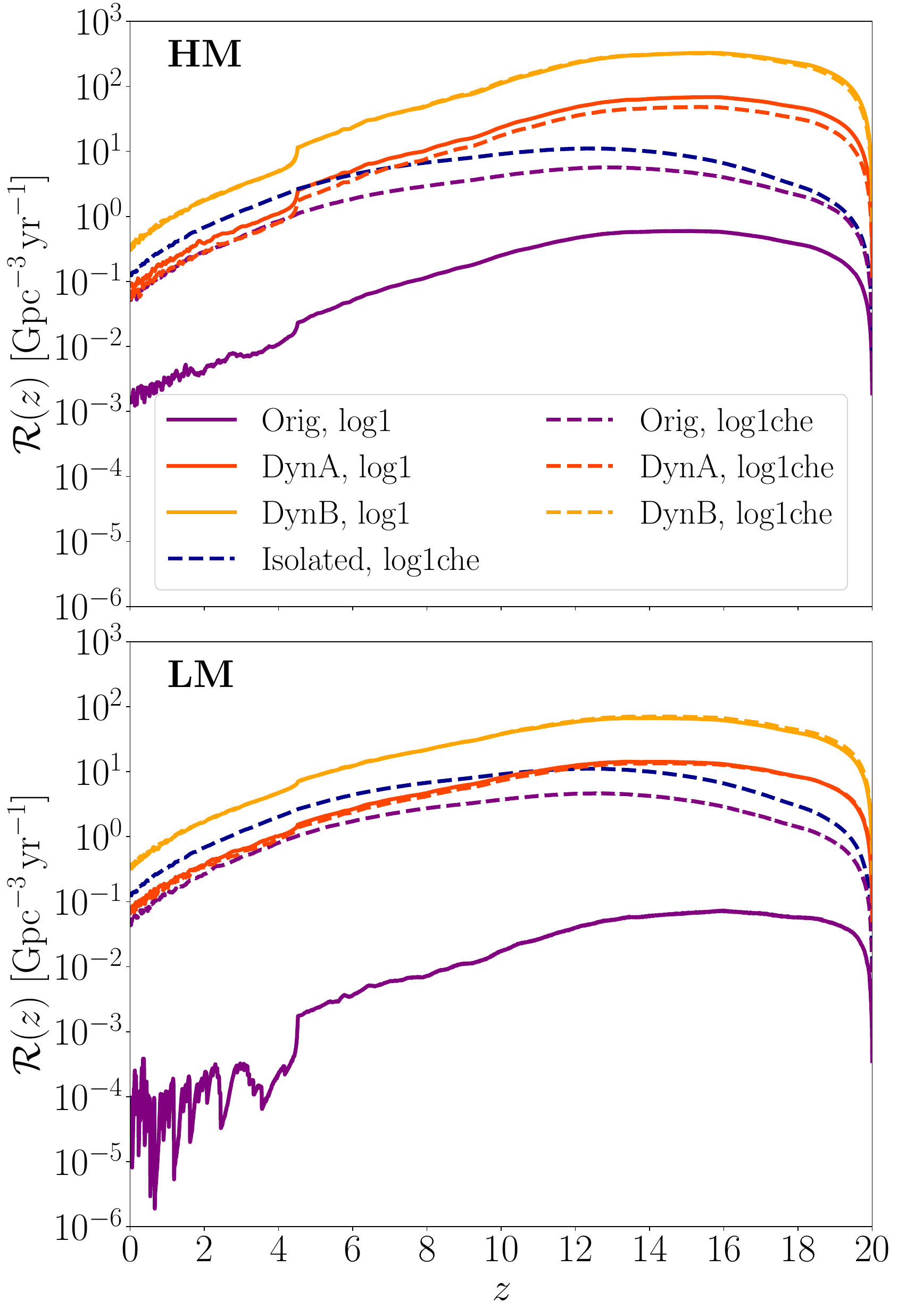}
    \caption{Same as Fig.~\ref{fig:mrd_gc_ysc} but only for BBH mergers with primary mass above the pair-instability mass gap. Isolated Pop.~III stars that do not evolve via CHE yield a merger rate density too low to be shown in the figure.} 
    \label{fig:mrd_gc_ysc_large}
\end{figure}

Figure~\ref{fig:mrd_gc_ysc_large} shows the merger rate density of BBHs with $m_1$ above the upper mass gap. In model DynB, the merger rate density of BBHs with primary mass above the gap is nearly the same as the total BBH merger rate density: nearly 100\% BBHs are above the mass gap in model DynB.

\section{Summary} \label{conclusions}
We have investigated the hierarchical mergers of Pop.~III binary black holes (BBHs) in dense star clusters, exploring a wide range of initial conditions for both the initial-mass functions (IMFs) and formation channels of first-generation BBHs. 

We find that star-cluster dynamics boosts the merger rate of Pop.~III BBHs with mass above the pair-instability mass gap by several orders of magnitude. Our models predict a dramatic difference between isolated and dynamical BBHs. Pop.~III BBH mergers above the gap are rare in isolation ($\lesssim{3}\%$), while they are extremely common in dynamical BBHs. This difference is maximal with model DynB, which always yields $\gtrsim{98}\%$ BBH mergers with primary mass $m_1 > 200\,\msun$. 

Models with chemically homogeneous evolution (CHE) show the least differences between isolated and dynamical BBHs above the gap, because CHE favors the merger of isolated BBHs above the mass gap compared to "default" stellar evolution. Even in this case, however, star-cluster dynamics allows the pair-up and merger of more massive BBHs and the formation of hierarchical chains that push $m_1$ up to $\sim 700\,\msun$.

Our star cluster models predict several sharp features in the $m_1-q$ distribution of BBH mergers.
We find a zone of avoidance in the $m_1-q$ plane, corresponding to the values for which either $m_1$ or $m_2$ are above the pair-instability mass gap. Such feature is not apparent for isolated Pop.~III BBH mergers because of the scarcity of mergers above the gap, unless we assume CHE. 

The initial properties of the simulated star clusters affect the number of generations of mergers and the mass distributions. In fact, because of their lower escape velocity, low-mass (LM) clusters host at most two generations of mergers and reach a maximum primary mass of $\sim1000\,\msun$. High-mass (HM) clusters instead, host at most three generations of mergers, and reach $m_1\sim1200\,\msun$.

Different initial conditions result in peculiar features of the mass-ratio distribution of original BBHs.  Models assuming initial orbital periods from \cite{sb2013} present a peak at $q\sim0.15$. These BBHs do not merge within a Hubble time in isolation \citep{costa2023}, while in star clusters their merger is triggered by dynamical hardening. Models assuming initial orbital periods from \cite{sana2012} show a peak at $q\sim0.8$, mainly due to BBHs formed through stable mass transfer. 

All our dynamical BBH models have a strong preference for nearly equal-mass systems. However, they also present a peak at $q\sim0.15$, corresponding to BBHs with primary mass above the gap and secondary mass below the gap. The DynB models also predict a steep drop at $q$ below 0.4--0.5 because of their strong preference for mergers with both $m_1$ and $m_2$ above the mass gap.

To take into account the effects of tidal disruption, we evolved our clusters up to a minimum redshift $z_{\rm min}$. We have shown that the $m_1-q$ distributions for $z_{\rm min}=10,\,6,\,2$ are almost identical. In fact, the most massive BBHs in the clusters merge by $z=10$, while lower mass BBHs are ejected by three-body interactions and merge outside the cluster by $z=0$.

In HM clusters, the merger rate density of dynamical BBHs is generally higher than the one of both original and isolated BBHs: it peaks at $z\sim15$ and rapidly decreases at lower redshifts. We find a maximum value  $\mathcal{R}(z=15)\approx{}200$ Gpc$^{-3}$ yr$^{-1}$. At redshift zero, $\mathcal{R}(z=0)\approx{}0.32$ Gpc$^{-3}$ yr$^{-1}$ is still a factor of $\sim{13}$ higher than the one of isolated and original BBHs, if we neglect CHE. The merger rate density of dynamical BBHs in LM clusters is lower than in HM clusters by up to one order of magnitude. 

Isolated and original BBHs born from CHE Pop.~III stars are quite exceptional also in terms of their merger rate: at redshift zero it is similar to the one of dynamical BBHs. The reason is that these stars evolve with very compact radii and produce tight BBHs that merge more efficiently.

In summary, our results indicate that, if Pop.~III stars form and evolve in dense star clusters, dynamical interactions 
crucially enhance the formation of Pop.~III BBH mergers  above the pair-instability mass gap ($\gtrsim{}200$~M$_\odot$) compared to isolated binary evolution. 

\begin{acknowledgements}
We thank Ralf Klessen, Veronika Lipatova, and Simon Glover for useful discussions. MM, ST, GC, GI, and FS acknowledge financial support from the European Research Council for the ERC Consolidator grant DEMOBLACK, under contract no. 770017. MM and ST also acknowledge financial support from the German Excellence Strategy via the Heidelberg Cluster of Excellence (EXC 2181 - 390900948) STRUCTURES. 
MB and MM also acknowledge support from the PRIN grant METE under the contract no. 2020KB33TP. FS acknowledges financial support from the AHEAD2020 project (grant agreement n. 871158). 
MAS acknowledges funding from the European Union’s Horizon 2020 research and innovation programme under the Marie Skłodowska-Curie grant agreement No.~101025436 (project GRACE-BH, PI: Manuel Arca Sedda). MAS acknowledges financial support from the MERAC Foundation. GC acknowledges financial support by the Agence Nationale de la Recherche grant POPSYCLE number ANR-19-CE31-0022.
\end{acknowledgements}

\bibliographystyle{aa} 
\bibliography{bibliography.bib} 

\begin{appendix}
\section{Initial conditions for Pop.~III BBHs}\label{ap:isolated}


In the following we describe the details of the initial conditions of our first-generation BHs, introduced in Sect. \ref{subsec:popIII}. We generated both the binary and single star simulations with the population synthesis code \textsc{sevn} \citep{sevn2023}. 
 We adopt the rapid model for supernova explosions \citep{fryer2012}, which enforces a mass gap between neutron stars and BHs. For pulsational instability supernovae, we consider the prescription presented in the Appendix of \cite{mapelli2020}. For binary evolution, we assume a common envelope parameter $\alpha_{\rm CE}\,=\,3$, while $\lambda_{\rm CE}$ is set self-consistently as in  \cite{claeys2014}. 
Here we report the initial distribution for the primary mass, mass ratio, orbital period and eccentricity of original binaries. All the cases and combinations adopted in this work are listed in Table \ref{table:ic_costa}.

\subsection{Original BBHs} \label{app:sec_orig}

Because of the lack of consensus on the IMF of Pop.~III stars, we consider a range of IMFs for the primary mass:
\begin{enumerate}[(i)]
    \setlength\itemsep{0.3em}
    \item A \cite{kroupa2001} IMF 
        $\xi(M_{\mathrm{ZAMS}})\propto M_{\mathrm{ZAMS}}^{-2.3}$,
    which is commonly adopted for Pop.~II stars. With respect to the canonical IMF, here we consider a single slope since $m_{\textrm{min}}\geq0.5\,\msun$.
    \item A \cite{larson1998} IMF  
        $\xi(M_{\mathrm{ZAMS}})\propto M_{\mathrm{ZAMS}}^{-2.35}\exp{\left(-M_{\mathrm{cut1}}/M_{\mathrm{ZAMS}}\right)}$,
    with $M_{\mathrm{cut1}}=20\,\msun$.
    \item A flat log distribution   $\xi(M_{\mathrm{ZAMS}})\propto M_{\mathrm{ZAMS}}^{-1}$ (\citealp{sb2013}; \citealp{hirano2015, susa2014, wollenberg2020, chon2021, tanikawa2021, jaura2022, prole2022}).
    We consider this IMF as the fiducial one in our work. 
    \item A top heavy distribution (\citealp{sb2013}; \citealp{jaacks2019, liu2020}), 
        $\xi(M_{\mathrm{ZAMS}})\propto M_{\mathrm{ZAMS}}^{-0.17}\exp{\left(-M_{\mathrm{cut2}}^2/M_{\mathrm{ZAMS}}^2\right)}$,
    where $M_{\mathrm{cut2}}=20\,M_{\odot}$.
\end{enumerate}

\subsubsection{Mass ratio and secondary mass}
We draw the secondary mass of the ZAMS star using three different prescriptions:
\begin{enumerate}[(i)]
    \setlength\itemsep{0.3em}
    \item The $q$ distribution from \cite{sana2012},
    $\xi(q)\propto q^{-0.1}$, with $q\in[0.1,1]$ and $M_{\mathrm{ZAMS,2}}\geq2.2\,\msun$,
    which fits to the mass ratio of O- and B-type binaries in the local Universe.
    \item A sorted distribution: we draw the ZAMS mass of the entire stellar population from the chosen IMF. Then, we randomly pair two stars, enforcing $M_{\mathrm{ZAMS,2}}\leq M_{\mathrm{ZAMS,1}}$. 
    \item The $q$ distribution from \citet{sb2013},
    $\xi(q)\propto q^{-0.55}$, with $q\in[0.1,1]$ and $M_{\mathrm{ZAMS,2}}\geq2.2\,\msun$.
    This distribution was obtained by fitting Pop.~III stars generated in cosmological simulations. 
\end{enumerate}

\subsubsection{Orbital period}
We adopt two distributions for the initial orbital period $P$.
\begin{enumerate}[(i)]
    \setlength\itemsep{0.3em}
    \item $\xi(\Pi)\propto \Pi^{-0.55}$, with $\Pi=\log{(P/\mathrm{day})}\in[0.15,5.5]$ from \cite{sana2012}. 
    \item  $\xi(\Pi)\propto \exp{\left[-\left(\Pi-\mu\right)^2/ \left(2\sigma^2\right)\right]}$, 
    that is a Gaussian distribution with $\mu=5.5$ and $\sigma=0.85$ \citep{sb2013}. 
\end{enumerate}

\subsubsection{Eccentricity}
We draw the orbital eccentricity $e$ from two distributions.
\begin{enumerate}[(i)]
    \setlength\itemsep{0.3em}
    \item $\xi(e)\propto e^{-0.42}$ with $e\in[0,1)$ \citep{sana2012}. 
    \item A thermal distribution $\xi(e)=2e$ with $e\in[0,1)$ \citep{heggie1975,tanikawa2021, hartwig2016, kinugawa2014}, 
    which favors highly eccentric systems. As shown by \cite{park2021,park2023}, Pop.~III binaries form preferentially with high orbital eccentricity. 
\end{enumerate}

\begin{figure}[ht]
    \centering
    \includegraphics[width=\columnwidth]{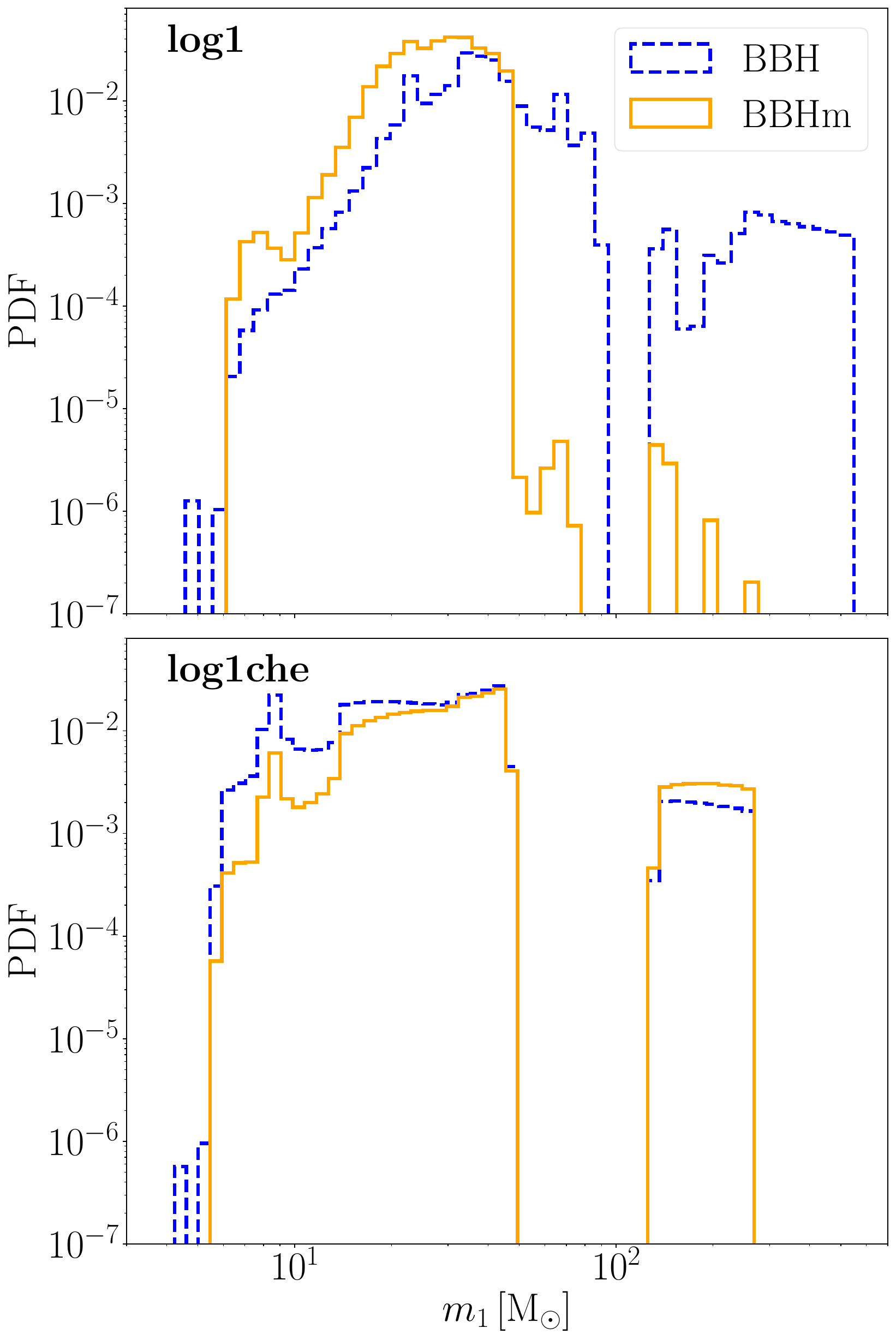}
    \caption{Distribution of BBHs (blue dashed line) and BBH mergers (orange solid line) from isolated binary star progenitors \citep{costa2023,santoliquido2023}. 
    We show the  \textbf{log1} (upper panel) and \textbf{log1che} (lower panel) models.}
    \label{fig:isolated}
\end{figure}

\begin{table*}[ht!]
\centering
\begin{tabular}{c c c c c c c c} 
 \hline
 Model & $M_{\mathrm{ZAMS,1}}$ & $M_{\mathrm{ZAMS}}$ & $q$ & $P$ & $e$ & $N\,(\times10^7)$ & $M_{\rm{tot}}\,(\times10^9\,\rm{M}_{\odot})$\\ [0.5ex] 
 \hline\hline
 \textbf{kro1} & \cite{kroupa2001} & Flat in log & \citetalias{sana2012} & \citetalias{sana2012} & \citetalias{sana2012} & 2 & 0.89\\ 
 \textbf{kro5} & \cite{kroupa2001} & Flat in log & \citetalias{sb2013} & \citetalias{sb2013} & Thermal & 2 & 0.93\\
 \textbf{lar1} & \cite{larson1998} & - & \citetalias{sana2012} & \citetalias{sana2012} & \citetalias{sana2012} & 2 & 1.2\\ 
 \textbf{lar5} & \cite{larson1998} & - & \citetalias{sb2013} & \citetalias{sb2013} & Thermal & 2 & 1.24\\
 \textbf{log1} & Flat in log & - & \citetalias{sana2012} & \citetalias{sana2012} & \citetalias{sana2012} & 1.45 & 2.59\\ 
 \textbf{log2} & Flat in log & - & \citetalias{sana2012} & \citetalias{sb2013} & Thermal & 1.45 & 2.58\\
 \textbf{log3} & - & - & Sorted & \citetalias{sana2012} & \citetalias{sana2012} & 1.38 & 3.19\\
 \textbf{log4} & - & - & \citetalias{sb2013} & \citetalias{sana2012} & Thermal & 1.53 & 2.6\\
 \textbf{log5} & Flat in log & - & \citetalias{sb2013} & \citetalias{sb2013} & Thermal & 1.53 & 2.6\\
 \textbf{top1} & Top heavy & - & \citetalias{sana2012} & \citetalias{sana2012} & \citetalias{sana2012} & 1.05 & 4.16\\ 
 \textbf{top5} & Top heavy & Flat in log & \citetalias{sb2013} & \citetalias{sb2013} & Thermal & 1.07 & 4.03\\
 \hline
\end{tabular}
\caption{Initial conditions assumed for the stellar models considered in this work. Column 1: name of the model. Column 2: IMF for the primary star. Column 3: ZAMS mass of the whole stellar population. Columns 4, 5 and 6: initial distributions for the mass ratio $q$, period $P$ and eccentricity $e$. The acronyms \citetalias{sana2012} and \citetalias{sb2013} refer to \cite{sana2012} and \cite{sb2013}, respectively. Column 7 and 9:  total number and total mass of the simulated binaries.}
\label{table:ic_costa}
\end{table*}




\subsection{Chemically-homogeneous stars}\label{ap:che}
In this work, we model CHE from pure-He stars, following \cite{santoliquido2023}. Fast-spinning metal-poor stars effectively become fully mixed by the end of the main sequence \citep{mandel2016,demink2016}. For this reason, pure-He stars can be used to describe the successive phases of their evolution \citep{tanikawa2021b}. 
In our treatment, we use pure-He stars as a proxy for CHE after the main sequence \citep{santoliquido2023}. To account for the hydrogen burning phase, we assume that the evolution of pure-He stars starts at the end of the main sequence in our dynamical simulations with \fastcluster. 


CHE stars remain pretty compact during their entire evolution \citep{tanikawa2021}. 
This generally prevents Pop.~III chemically-homogeneous stars from merging in the early stages of their life. 
Figure~\ref{fig:isolated} compares the distributions of isolated Pop.~III BBHs and BBH mergers from models \textbf{log1} and \textbf{log1che} \citep{costa2023,santoliquido2023}. Isolated BBH mergers with primary mass $m_{1}>10^2\,\msun$ are extremely rare, because their stellar progenitors merge with their companion when they become giant stars. In contrast, CHE stars evolve with compact radii for their entire life, thus preventing the BH progenitors from becoming giant stars at the end of the main sequence. As a consequence, CHE binaries can efficiently produce BBH mergers above the upper mass gap.

\section{Pairing functions}\label{ap:pairing}
We draw the masses of our dynamical BBHs from catalogs of single BHs obtained with \textsc{sevn} \citep{sevn2023}. We consider the same initial conditions as for original binaries (Sect. \ref{app:sec_orig}). 
Since the $P$ and $e$ distributions are set based on dynamical prescriptions (eq. \ref{eq:ecc_peri}), we only draw the BH mass from the \textsc{sevn} catalogs. Thus, the only parameters of interest for dynamical BBHs in Table \ref{table:ic_costa} are $M_{\mathrm{ZAMS,1}}$, $N$ and $M_{\rm{tot}}$. 
We draw these parameters from the models \textbf{kro1}, \textbf{lar1}, \textbf{log1} and \textbf{top1}.

We consider two different coupling criteria for generating the first-generation dynamical BBH masses.
In model DynA, we sample the mass of the primary BH uniformly from the \textsc{sevn} single BH catalog \citep{mapelli2021}, while the secondary is drawn from the probability distribution function \citep{oleary2016}:

\begin{equation}
    p(m_2) \propto (m_1+m_2)^4\,\,\,\,\,\,\,\,\mathrm{with}\,\,m_2\in[5\,\msun,m_1).
\end{equation}

\begin{figure}[ht]
    \centering
    \includegraphics[width=\columnwidth]{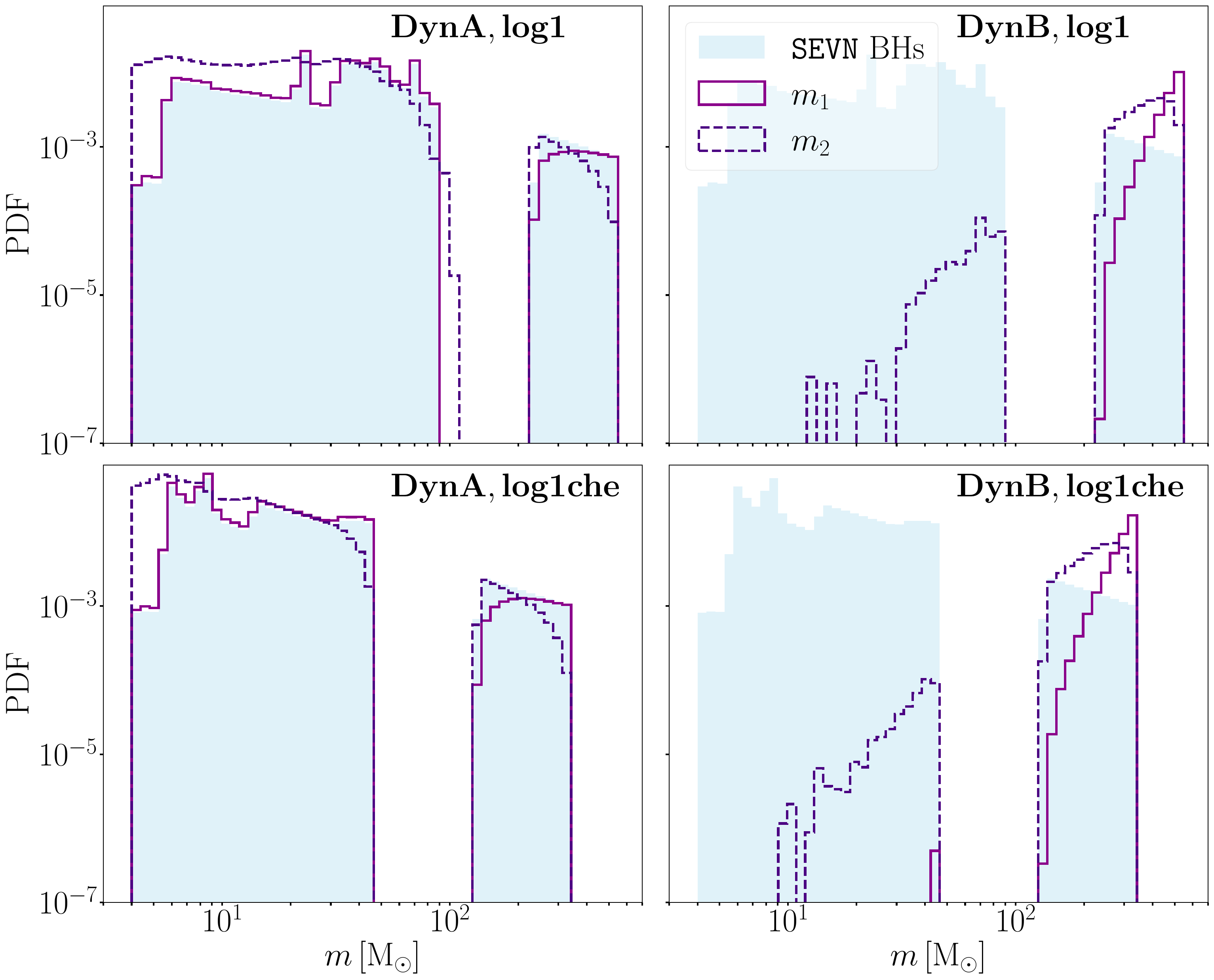}
    \caption{Distribution of $m_1$ (purple solid line) and $m_2$ (violet dashed line) for first-generation BBHs, for model DynA (left) and DynB (right). The blue shaded area represents the distribution of single BHs from the \textsc{sevn} catalog.  We show the  \textbf{log1} (upper panel) and \textbf{log1che} (lower panel) models. }
    \label{fig:sampling}
\end{figure}

In model DynB, we derive the primary and secondary BH masses following the sampling criterion introduced in \cite{antonini2023}, which is based on the formation rate of hard binaries per unit of volume and energy  \cite{heggie1975}:

\begin{equation}\label{eq:heggie}
    \Gamma_{\rm 3b}(m_1,m_2,m_3)\propto\frac{n_1\,n_2\,n_3\,m_1^4m_2^4m_3^{5/2}}{(m_1+m_2+m_3)^{1/2}(m_1+m_2)^{1/2}}\,\beta^{9/2},
\end{equation}

with $n_i=n(m_i)$ the number density of BHs with mass $m_i$ ($i$=1,2,3) and $\beta$ the inverse of their mean internal energy. For each model, we derive $n(m)$ from the mass function of single BHs in the \textsc{sevn} catalog. Since eq.~\ref{eq:heggie} is symmetric in $m_1$ and $m_2$, the probability density function for either of them is

\begin{equation}\label{eq:h75}
    p_{1,2}(m_{1,2})=\int_{m_{\rm low}}^{m_{\rm up}}\int_{m_{\rm low}}^{m_{\rm up}}dm_3\,dm_{2,1}\,\Gamma_{\rm 3b}(m_1,m_2,m_3),
\end{equation}

where $m_{\rm low}$ and $m_{\rm up}$ are the lower and upper limit of the mass distribution. Using eq. \ref{eq:h75}, we sample the primary and secondary mass of the BBHs and then we assign the largest one to $m_1$ and the smallest to $m_2$.
As shown in Fig.~\ref{fig:sampling}, the model DynA tends to reproduce the underlying BH mass function, while DynB favors the coupling of the most massive BHs. 

\section{Uncertainties related to stochastic fluctuations on model DynB}
The sampling criterion in model DynB strongly favors the coupling of the most massive BHs, as expected from dynamical interactions in high-density star clusters \citep{heggie1975}. However, only a limited number of BHs may be present in a single star cluster, and in particular in those with  $M_{\rm cl}\lesssim 10^4 \, \msun$. 

Here, we quantify the effect of stochastic fluctuations due to the limited number of BHs in a star cluster on the BBH mergers predicted by model DynB. 
First, we define the formation rate of hard binaries per unit of volume and energy (eq. \ref{eq:heggie}), as described in Appendix \ref{ap:pairing}. Then, we randomly draw a sub-sample with size $N$ from the \textsc{sevn} catalog, to mimic the presence of a certain number of BHs within the cluster. Finally, we sample the BH masses from this set, instead of the whole \textsc{sevn} catalog, based on the probability density function defined in eq. \ref{eq:h75}. 

\begin{figure}[ht]
    \centering
    \includegraphics[width=\columnwidth]{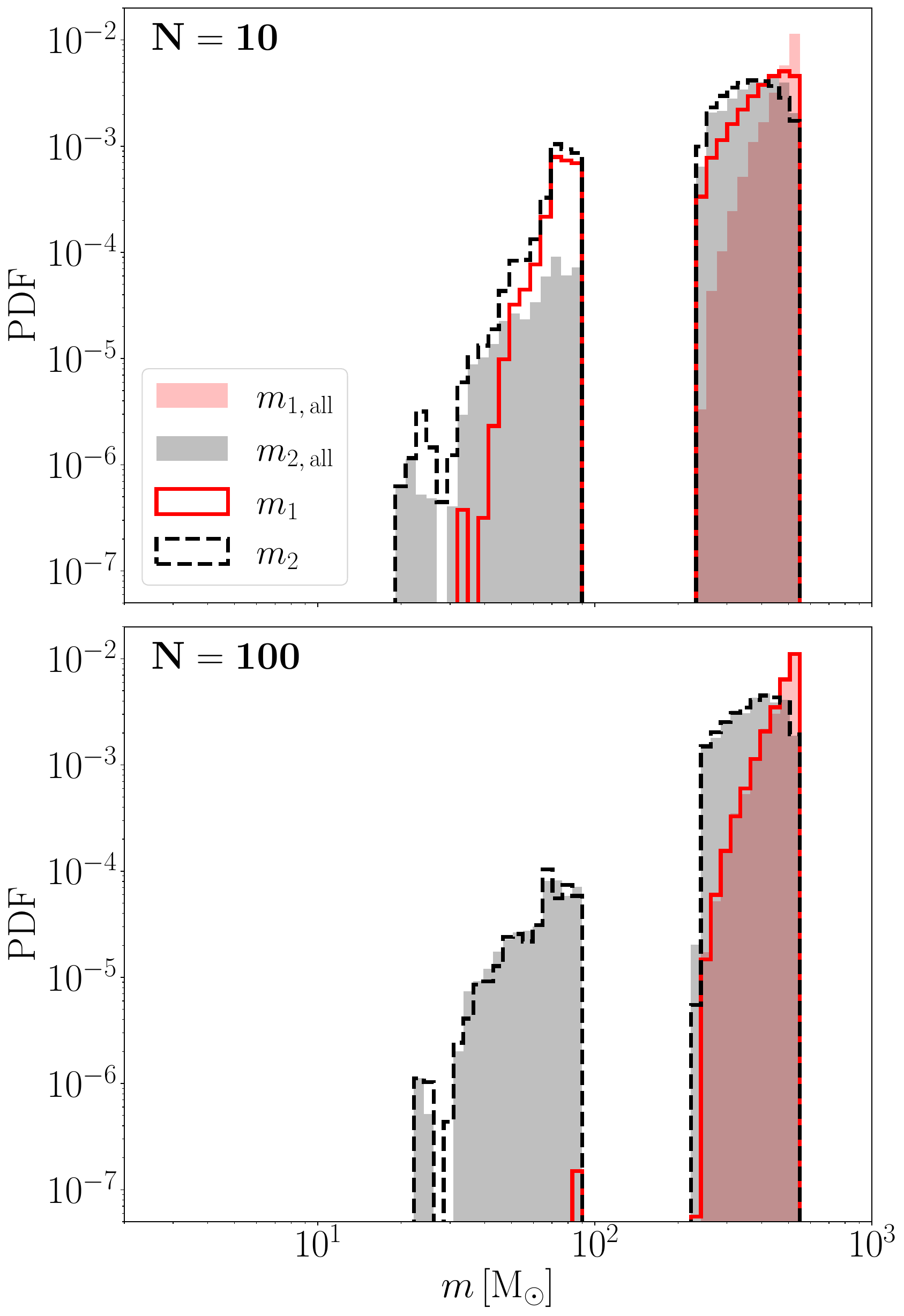}
    \caption{Distribution of $m_1$ (red solid line) and $m_2$ (black dashed line) for BBH mergers with model DynB in HM clusters. We obtain them considering, at each drawing, sub-samples of the \textsc{sevn} catalog with size 10 (upper panel) and 100 (lower panel), respectively. The shaded areas show the distributions obtained when sampling the 1g BHs from the \textsc{sevn} catalog without any restriction.}
    \label{fig:gc_ncfr}
\end{figure}

Figure \ref{fig:gc_ncfr} shows the primary and secondary BH masses in BBH mergers, obtained by considering $N=10$ and $100$ BHs for model \textbf{log1}. The BH distributions for $N=100$ are almost identical to the case without restrictions, because the sub-samples are large enough to contain a number of massive BHs, which are most likely coupled in our model. In contrast, when $N=10$ the very limited number of BHs allow for the coupling of less massive BHs, with primary mass $m_1 \lesssim 100 \, \msun$. As a consequence, the primary mass distribution of BBH mergers now extends below the pair-instability mass gap, down to $\approx 20 \, \msun$.

\section{Effects of tidal disruption} \label{ap:tidal_gc}
As discussed in Sec.~\ref{subsec:disruption_gc}, we simulated the evolution of our clusters up to a minimum redshift $z_{\rm min}$, with $z_{\rm min}=2$, $6$ and $10$ and evaluate the impact of $z_{\rm min}$ on the number of BBH mergers. In the following, we account for the effects of tidal disruption on the number and mass distributions of BBH mergers.

\begin{figure*}[ht]
    \centering
    \includegraphics[width=\textwidth]{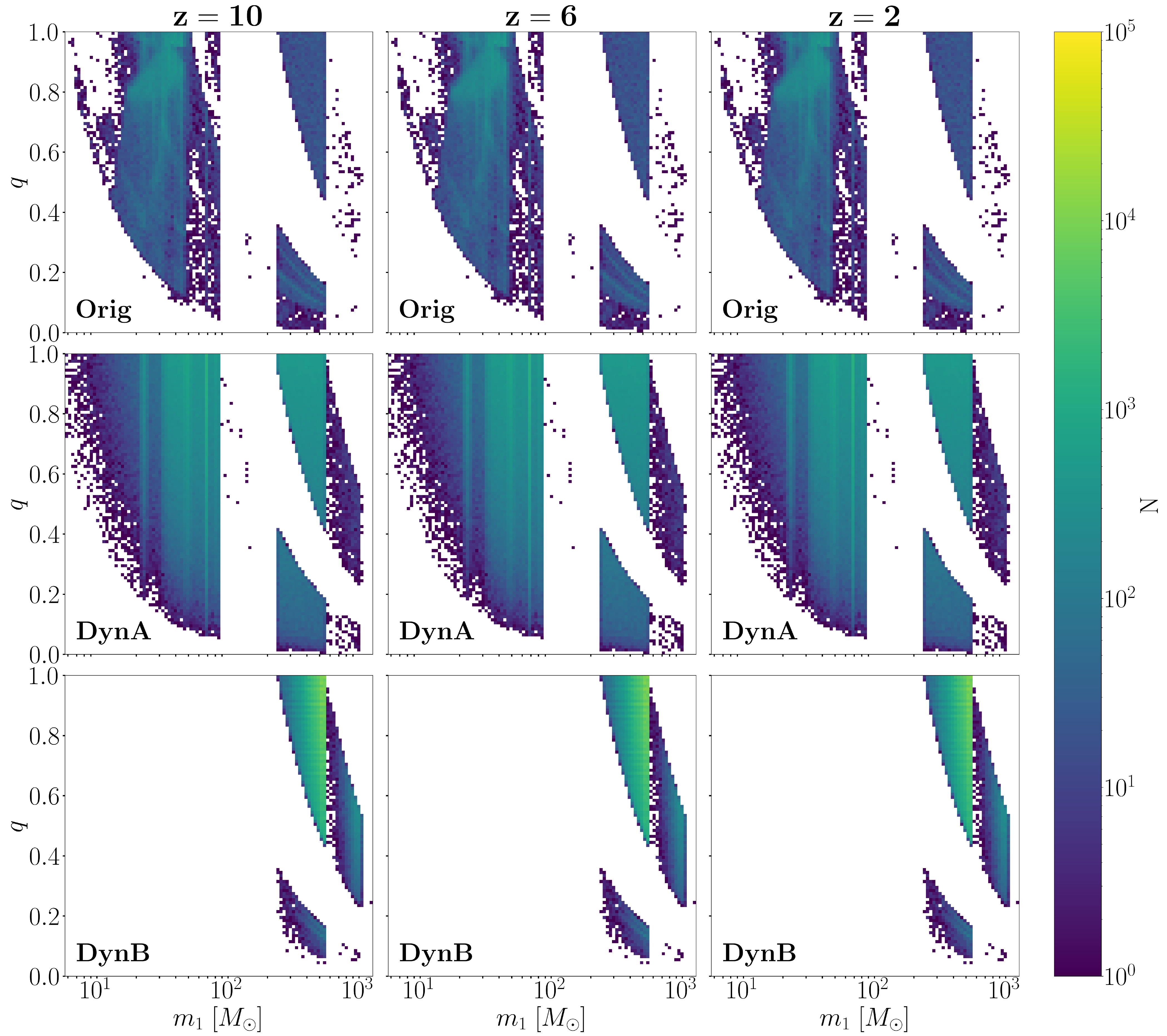}
    \caption{
Distribution of the mass ratio $q$ as a function of the primary mass $m_1$ for model \textbf{log1}, assuming that the HM cluster gets disrupted at  redshift $z_{\rm min}=10$, 6, 2. We show the results both for original (Orig) and dynamical (DynA and DynB) binaries.}
    \label{fig:m1q_z_log1}
\end{figure*}

We find negligible difference in the populations of BBH mergers at different $z_{\rm min}$. Table~\ref{table:residuals_z} reports the percentage variation of the number of BBH mergers between $z_{\rm min}=10$ and $z_{\rm min}=2$. Such variation is always smaller than $0.05\%$. 
Fig.~\ref{fig:m1q_z_log1} shows the distributions of BBH mergers in the $m_1-q$ plane, for different $z_{\rm{min}}$.
Independently of the model considered, there are minimal differences between the mass spectra of BBH mergers at $z_{\rm min} \leq 10$.

\begin{table}[ht!]
\centering
\begin{tabular}{c c c c} 
\hline
IC & Orig & DynA & DynB\\ [0.5ex] 
& \% & \% & \%\\
\hline\hline
\textbf{log1} & 0.005 & 0.005 & 0.028 \\
\textbf{log1che} & 0.011 & 0.013 & 0.051 \\
\hline
\end{tabular}
\caption{Percentage variation in the number of BBH mergers between $z_{\rm min}=10$ and $z_{\rm min}=2$. Column 1: name of the model; columns 2, 3, 4: BBH formation channel.}
\label{table:residuals_z}
\end{table}

\end{appendix}
%
%

\end{document}